\documentclass[reprint,amsmath,amssymb,aps,pra,longbibliography]{revtex4-2}

\usepackage{graphicx}
\usepackage{dcolumn}
\usepackage{bm}
\usepackage{xcolor}
\usepackage{mathptmx}
\usepackage{comment}

\begin{document}

\title{Limits on the free-space group velocity of optical wave packets incorporating angular dispersion. Part~I, conventional angular dispersion: tutorial}

\author{Layton A. Hall$^{1,2}$}
\author{Ayman F. Abouraddy$^1$}
\email{raddy@creol.ucf.edu}
\affiliation{$^1$CREOL, The College of Optics \& Photonics, University of Central Florida, Orlando, FL 32816, USA}
\affiliation{$^2$Los Alamos National Laboratory, Los Alamos, NM 87545, USA}

\begin{abstract}
A plane-wave optical pulse travels in free space at a group velocity $c$ (the speed of light in vacuum) measured along its propagation axis. It is sometimes thought that spatially structuring the field in free space reduces the group velocity below $c$ because -- intuitively -- the oblique wave vectors undergirding the wave packet increase the average group delay. Here we show that spatiotemporally structuring a pulsed optical beam (or wave packet) can yield -- in principle -- arbitrary group velocities in free space in contradistinction to this commonly held intuition. We devote our attention here to pulses endowed with angular dispersion (AD) where the propagation angle is wavelength dependent. This class of wave packets is particularly pertinent because their group velocity is constant over the transverse field profile and along the propagation axis, thereby yielding an unambiguous group delay. However, significant deviation of the AD-induced group velocity in free space from~$c$ inevitably requires a large numerical aperture that lies deep in the non-paraxial regime, and such wave packets experience AD-induced group-velocity dispersion. The formulation presented here captures a broad range of results, connecting them in a single framework. In Part~II of this tutorial, we describe recently identified `non-differentiable AD' (associated with propagation-invariant space-time wave packets) that helps circumvent the limits associated with conventional (differentiable) AD as outlined here.
\end{abstract}

\maketitle

\section{Introduction}

\begin{figure}[t!]
\centering
\includegraphics[width=8cm]{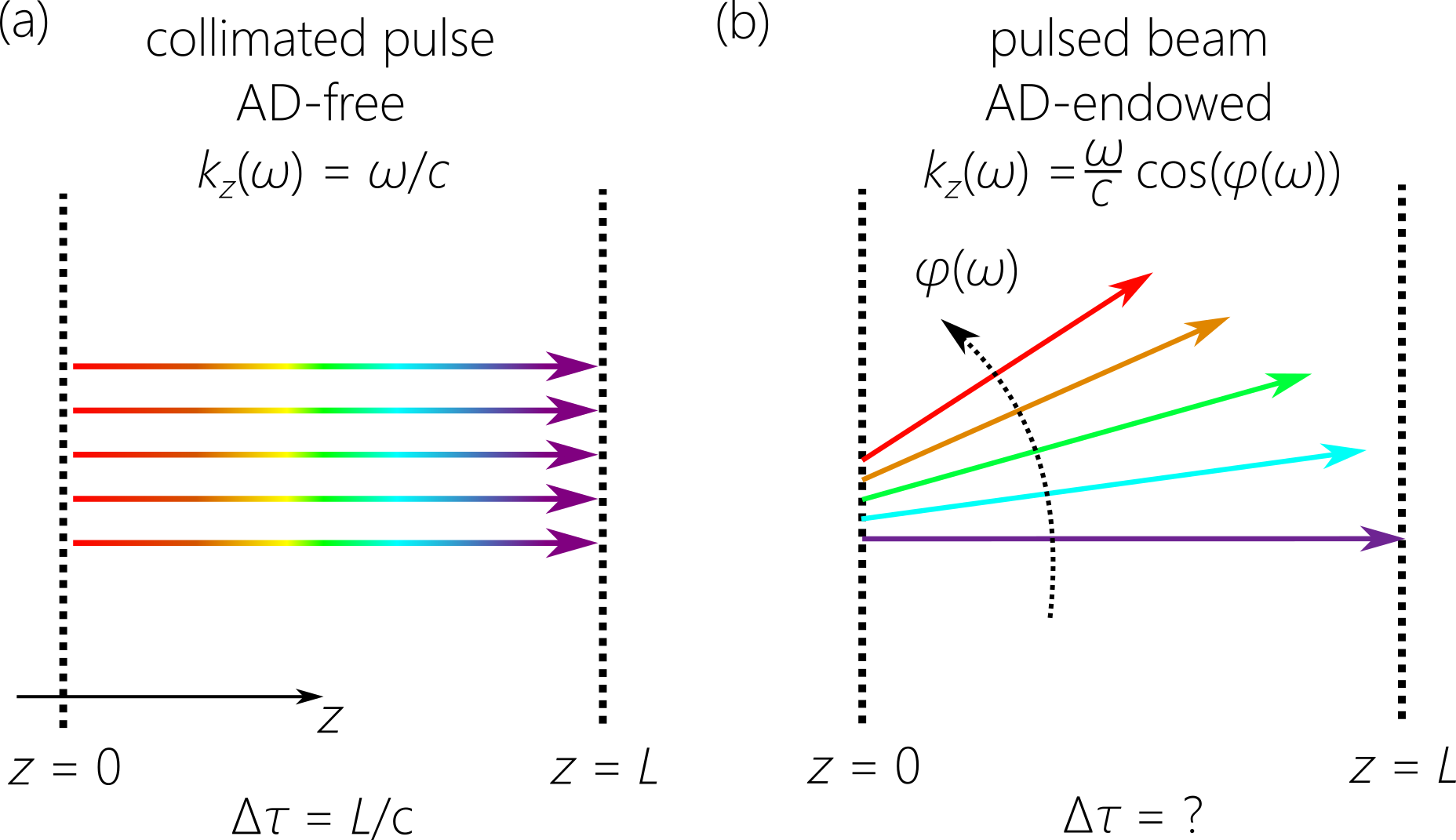}
\caption{Does spatial structuring of an optical pulse always reduce its group velocity? (a) A collimated, AD-free pulse in which all the frequencies $\omega$ travel in the same direction along the $z$-axis. The group delay when traveling between the planes $z=0$ to $z=L$ is $\Delta\tau=\tfrac{L}{c}$, and the group velocity along the $z$-axis is $c$. (b) A structured pulse in which AD is introduced. Each temporal frequency $\omega$ travels at a different angle $\varphi(\omega)$ with the $z$-axis. Intuitively, the effective group delay $\Delta\tau$ is expected to be larger than that in (a), corresponding to a drop in group velocity below~$c$. In this tutorial we show that this intuitive picture does not hold true in all scenarios.}
\label{fig:Question}
\end{figure}

The group velocity $\widetilde{v}$ of a spatially collimated optical pulse measured along an axis normal to its phase front in free space is $c$ (the speed of light in vacuum); see Fig.~\ref{fig:Question}(a). What is $\widetilde{v}$ when spatial or spatiotemporal structure is introduced into the field that yields a pulsed \textit{beam} [Fig.~\ref{fig:Question}(b)]? Intuitively, the group velocity is expected to drop below $c$ upon such structuring \cite{Sambles15Science}. The rationale is straightforward: the pulsed \textit{beam} (or wave packet) comprises plane-wave components that travel obliquely to the propagation axis, and thus traverse a longer distance than their on-axis counterpart. This geometric picture implies an increase in the group delay $\Delta\tau$ incurred by the wave packet between two planes, corresponding to a drop in the group velocity, $\widetilde{v}<c$. An example of this is demonstrated in Ref.~\cite{Giovannini15Science}, where introducing a Bessel-beam spatial structure into a collimated field produced a \textit{subluminal} group velocity in free space. Nevertheless, only a minute deviation from $c$ was observed ($\widetilde{v}\approx0.99999c$), manifesting in a relative group delay of $\Delta\tau\approx33$~fs with respect to a collimated reference field, corresponding to a relative distance of $\approx 10$~$\mu$m after propagation for 1~m. From this \cite{Giovannini15Science} and subsequent \cite{Bouchard16Optica,Lyons18Optica} results, it is sometimes thought that spatial structuring always reduces the wave packet group velocity \cite{Sambles15Science}.

One of the aims of this paper is to dispel this common notion and show that the intuitive geometric picture depicted in Fig.~\ref{fig:Question} is indeed misleading: the group velocity of a spatiotemporally structured pulsed beam in free space may indeed be subluminal ($\widetilde{v}<c$), superluminal ($\widetilde{v}>c$), or even negative ($\widetilde{v}<0$). Although the basic theoretical underpinning of this fact has been known for a few decades \cite{Porras03PRE2}, this knowledge is not widely spread, mainly because only minute deviations $|\widetilde{v}-c|\sim10^{-5}c$ are realizable in the paraxial regime, thus requiring precise measurements for their observation \cite{Bowlan09OL,Giovannini15Science}. Indeed, significant deviation in $\widetilde{v}$ from $c$ necessitates either exorbitant resources \cite{Zapata06OL} or operation in the non-paraxial regime \cite{Porras03PRE}. Nevertheless, these small deviations are useful in phase-matching of nonlinear interactions \cite{Danielius96OL,Averchi08PRA,Hebling02OE,Hebling08JOSAB,Wang20LPR}.

Recent developments in spatiotemporally structured fields \cite{Yessenov22AOP,Shen23JO,Abouraddy25OPN} now make possible the realization of dramatic deviations in the group velocity from $c$ in free space \textit{while remaining in the paraxial regime} \cite{Kondakci16OE,Parker16OE,Kondakci17NP,SaintMarie17Optica,Froula18NP,Kondakci19NC}. It is thus important to elucidate why these recent experiments have made such effects possible when prior experiments could not observe them. Besides its interest at the fundamental level, control over the group velocity of a pulsed beam is useful in a broad variety of contexts, from phase-matching in nonlinear interactions between fields at different wavelengths \cite{Danielius96OL,Averchi08PRA,Hebling02OE,Hebling08JOSAB,Wang20LPR}, to optical delay lines \cite{Zapata08JOSAA,Alfano2016OC,Saari2017OC,Yessenov20NC}, laser-plasma interactions, electron acceleration, and dephasingless wakefield accelerators \cite{Vieira18PRL,Caizergues20NP,Palastro20PRL,Sun24PRR,Piccardo25Optica,Longman26NP,Vaz26arxiv}.

Of course, there is an extensive literature on `slow' and `fast' light where the group velocity of a pulse is tuned by exploiting active and absorptive resonances in optical materials or photonic structures \cite{Hau99Nature,Kash99PRL,Song05OE,Vlasov05N,Okawachi05PRL,Krauss07JPD,Krauss08NP,Baba08NP,Boyd09Science,Tsakmakidis17Science}. Such intrinsically narrow resonances restrict the group-velocity tunability to long pulses (narrow spectra) \cite{Parra07OPN}, and their delay-bandwidth product is limited to approximately unity (i.e., the group delay achievable is on the order of the pulse width) \cite{Tucker05EL,Tucker05JLT,Kurgin08Book} -- except when active interventions are implemented \cite{Yanik04PRL} (e.g., trapdoor scenarios \cite{Xu06NPhys,Tsakmakidis17Science2}). By contrast, spatiotemporally structuring a pulsed beam can allow for tuning its group velocity in free space \cite{Kondakci19NC,Yessenov19OE} and in optical media \cite{Bhaduri19Optica}, it enables utilization of broad bandwidths (ultrashort pulses) \cite{Kondakci18OE}, and it is not restricted by the delay-bandwidth-product theorem \cite{Yessenov20NC}.

The space of spatiotemporally structured optical fields is vast \cite{Yessenov22AOP,Shen23JO,Abouraddy25OPN}. However, from the perspective of tuning a wave-packet group velocity, a particularly useful sub-class is that of optical fields endowed with angular dispersion (AD) \cite{Torres10AOP,Fulop10Review}, whereupon each temporal frequency $\omega$ travels at a different angle $\varphi(\omega)$ with respect to a fixed axis [Fig.~\ref{fig:Question}(b)], a phenomenon that routinely occurs when a collimated polychromatic field traverses diffractive or dispersive devices (e.g., prisms, gratings, etc.). The pulsed Bessel beam in Ref.~\cite{Giovannini15Science} is indeed an example of a field endowed with AD. The AD-based model we focus on here has important advantages: the ensuing group velocity is in principle constant along the propagation axis \textit{and} over the transverse spatial cross section of the beam. Other structured fields (for example, pulsed vortex beams endowed with orbital angular momentum \cite{Bouchard16Optica} or tightly focused pulses \cite{Porras03PRE}) have group velocities that change along the propagation axis and/or vary across the wave packet transverse cross section (as pointed out in \cite{Saari17Optica}), which can make estimating the wave-packet group delay (and hence its group velocity) challenging. These difficulties are absent from AD-endowed wave packets, which makes them particularly useful in studying the tunability of $\widetilde{v}$ in free space.

We aim to achieve several goals in Part~I of this two-part tutorial (see the accompanying paper \cite{Hall26JOSAA2}). First, we dispel the notion that introducing spatial structure into an optical field necessarily reduces the group velocity below $c$ in free space; rather, AD-endowed wave packets can -- in principle -- be subluminal ($\widetilde{v}<c$), superluminal ($\widetilde{v}>c$), or even negative ($\widetilde{v}<0$) in free space. This is achieved without recourse to resonant optical media \cite{Kurgin08Book,Boyd09Science}, so there is no attenuation or amplification with propagation. This counter-intuitive result stems from a confluence of \textit{geometric} and \textit{interferometric} factors whose combination is yet to be fully appreciated by the broad optics and photonics community. However, significant deviation in $\widetilde{v}$ can\textit{not} be realized in the \textit{paraxial} regime. Part~I of this tutorial serves as a prelude to introducing the recently identified form of AD referred to as `non-differentiable' AD \cite{Hall21OL,Hall22OEConsequences,Hall22JOSAA,Hall25APLP}, which is associated with the family of pulsed beams known as space-time wave packets (STWPs) \cite{Yessenov22AOP}. We show in Part~II of this tutorial \cite{Hall26JOSAA2} that non-differentiable AD helps circumvent the rules-of-thumb outlined here for conventional (differentiable) AD.

This paper is structured as follows. First, we define AD through a perturbative expansion of the propagation angle $\varphi(\omega)$ with respect to the frequency $\omega$. This allows us to derive a general formula for $\widetilde{v}$ in free space for an AD-endowed wave packet. We then explore the various domains of $\widetilde{v}$: luminal, subluminal, superluminal, and negative. Moreover, we describe the AD-induced GVD incurred by the field in free space. We apply these results to two classes of fields: X-waves (which are superluminal) \cite{Saari97PRL} and pulsed Bessel beams (which are subluminal) \cite{Giovannini15Science}. In both cases, we show that only minute deviations in $\widetilde{v}$ from $c$ can be observed in the paraxial domain.

\section{Angular dispersion}

\subsection{Definition}

\begin{figure}[t!]
\centering
\includegraphics[width=8.6cm]{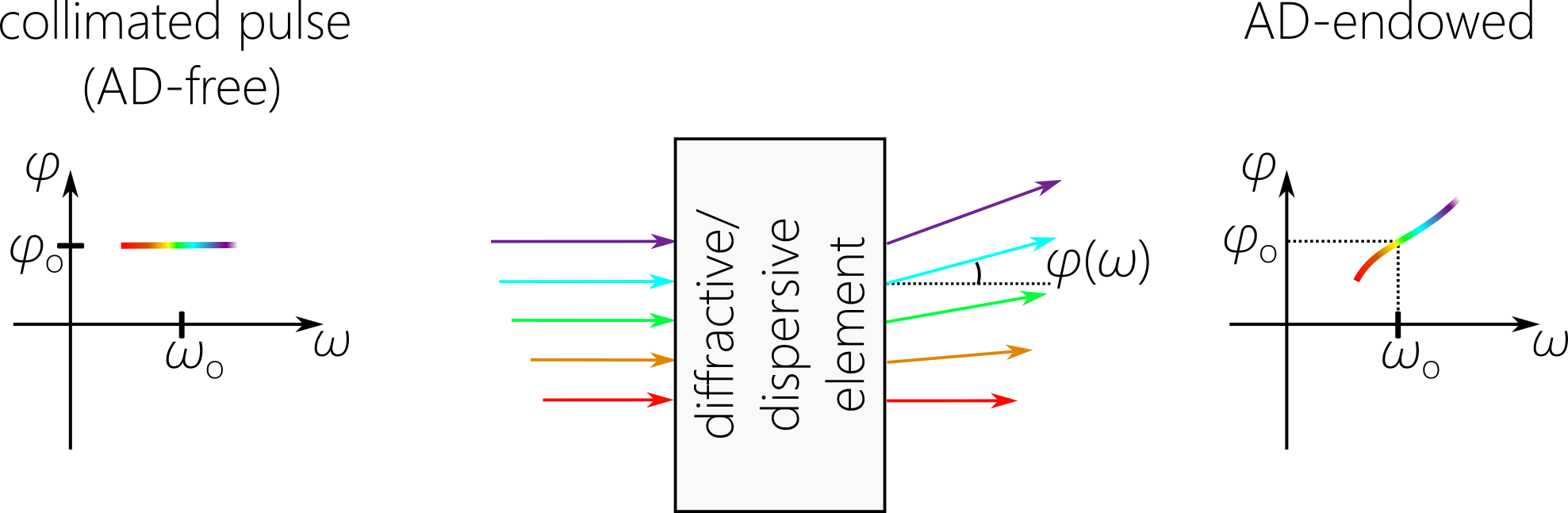}
\caption{Concept of angular dispersion (AD). A collimated pulse, in which all the frequencies $\omega$ travel at the same angle $\varphi(\omega)=\varphi_{\mathrm{o}}$ with the $z$-axis, traverses a diffractive or dispersive device that introduces AD, after which each frequency travels at a different angle $\varphi(\omega)$.}
\label{fig:Concept}
\end{figure}

Consider a collimated field traversing a dispersive or diffractive device after which each temporal frequency $\omega$ travels at a different angle $\varphi(\omega)$ to the optical axis, which we take to be the $z$-axis here [Fig.~\ref{fig:Concept}]. For conventional AD, the first step is to expand $\varphi(\omega)$ in a Taylor series around a fixed frequency $\omega_{\mathrm{o}}$ \cite{Porras03PRE2,Hall25APLP}:
\begin{equation}\label{eq:ADexpansion}
\varphi(\omega)=\varphi(\omega_{\mathrm{o}}+\Omega)\approx\varphi_{\mathrm{o}}+\varphi_{\mathrm{o}}^{(1)}\Omega+\frac{1}{2}\varphi_{\mathrm{o}}^{(2)}\Omega^{2}+\cdots,
\end{equation}
where $\Omega=\omega-\omega_{\mathrm{o}}$ is the frequency displacement from $\omega_{\mathrm{o}}$,  $\varphi_{\mathrm{o}}\!=\!\varphi(\omega_{\mathrm{o}})$ is the propagation angle at $\omega_{\mathrm{o}}$, the first-order coefficient $\varphi_{\mathrm{o}}^{(1)}=\tfrac{d\varphi}{d\omega}\bigr|_{\omega_{\mathrm{o}}}$ is usually known as AD for brevity, and $\varphi_{\mathrm{o}}^{(2)}\!=\!\tfrac{d^{2}\varphi}{d\omega^{2}}\big|_{\omega_{\mathrm{o}}}$. The contribution of each term in this perturbative expansion generally declines with increasing order. We refer to field configurations in which $\varphi_{\mathrm{o}}=0$ as `on-axis' fields, and `off-axis' fields otherwise when $\varphi_{\mathrm{o}}\neq0$ (we discuss this distinction in more detail below). Therefore, the propagation axis of an on-axis field coincides with the $z$-axis, while limited propagation distances can be utilized for off-axis fields unless $\varphi_{\mathrm{o}}$ is small.

The axial wave number $k_{z}(\omega)=\tfrac{\omega}{c}\cos\{\varphi(\omega)\}$ in free space can thus be expanded as:
\begin{eqnarray}\label{eq:AxialWaveNumber}
k_{z}(\omega)&\approx&\frac{\omega}{c}\cos\left(\varphi_{\mathrm{o}}+\varphi_{\mathrm{o}}^{(1)}\Omega+\frac{1}{2}\varphi_{\mathrm{o}}^{(2)}\Omega^{2}+\cdots\right)\nonumber\\
&\approx&k_{\mathrm{o}}\cos\varphi_{\mathrm{o}}+\frac{1}{\widetilde{v}}\Omega+\frac{1}{2}k_{2}\Omega^{2}+\cdots;
\end{eqnarray}
where $k_{\mathrm{o}}=\omega_{\mathrm{o}}/c$ is the free-space wave number at $\omega_{\mathrm{o}}$, $\widetilde{v}$ is the group velocity, and $k_{2}$ is the AD-induced GVD coefficient. This expansion is \textit{not} a small-angle approximation (indeed, $\varphi_{\mathrm{o}}$ can be large); rather, it is simply an expansion around $\omega_{\mathrm{o}}$. That is, the angular variation around $\varphi_{\mathrm{o}}$ is assumed to be small with respect to $\varphi_{\mathrm{o}}$ (unless $\varphi_{\mathrm{o}}=0$ for on-axis fields), and that the contribution of higher-order terms generally diminish. We take the lowest-order terms $\varphi_{\mathrm{o}}$, $\varphi_{\mathrm{o}}^{(1)}$, and $\varphi_{\mathrm{o}}^{(2)}$ to be the most significant. 

We can write a field incorporating AD as $E(x,z;t)=e^{ik_{\mathrm{o}}(z\cos\varphi_{\mathrm{o}}-ct)}\psi(x,z;t)$, where we hold the field uniform along $y$, and the slowly varying spatiotemporal envelope $\psi(x,z;t)$ is:
\begin{equation}
\psi(x,z;t)=\int\!d\Omega\;\widetilde{\psi}(\Omega)e^{i\{k_{x}(\Omega)x+[k_{z}(\Omega)-k_{\mathrm{o}}\cos\varphi_{\mathrm{o}}]z-\Omega t\}};
\end{equation}
here $\widetilde{\psi}(\Omega)$ is the complex spectral amplitude, which is the Fourier transform of $\psi(0,0;t)$. Note that this spectrum is defined by a single integral over $\Omega$, with both $k_{x}(\omega)=\tfrac{\omega}{c}\sin\varphi(\omega)$ and $k_{z}(\omega)=\tfrac{\omega}{c}\cos\varphi(\omega)$ depending on $\Omega$. Substituting for $k_{z}(\omega)$ from Eq.~\ref{eq:AxialWaveNumber} we have:
\begin{equation}
\psi(x,z;t)=\int\!d\Omega\;\widetilde{\psi}(\Omega)e^{i\{k_{x}(\Omega)x-\Omega(t-z/\widetilde{v})+\tfrac{1}{2}k_{2}\Omega^{2}z\}};
\end{equation}
that is, the envelope travels in free space at a fixed group velocity $\widetilde{v}$ that is the same across the entire transverse spatial profile and along $z$. However, the wave packet does \textit{not} travel rigidly at $\widetilde{v}$ (i.e., it is \textit{not} propagation invariant); rather, the wave packet propagation is accompanied by diffractive spatial spreading due to the $e^{ik_{x}(\Omega)x}$ term \textit{and} dispersive temporal spreading due to the $e^{i\tfrac{1}{2}k_{2}\Omega^{2}z}$ term. Note, of course, that the two terms $e^{ik_{x}(\Omega)x}$ and $e^{i\tfrac{1}{2}k_{2}\Omega^{2}z}$ are related through their co-dependence on $\Omega$ (which was exploited in producing the space-time Talbot effect \cite{Hall21APLSTTalbot}).

\subsection{Depiction of a wave packet in the AD plane}

\begin{figure}[t!]
\centering
\includegraphics[width=8.6cm]{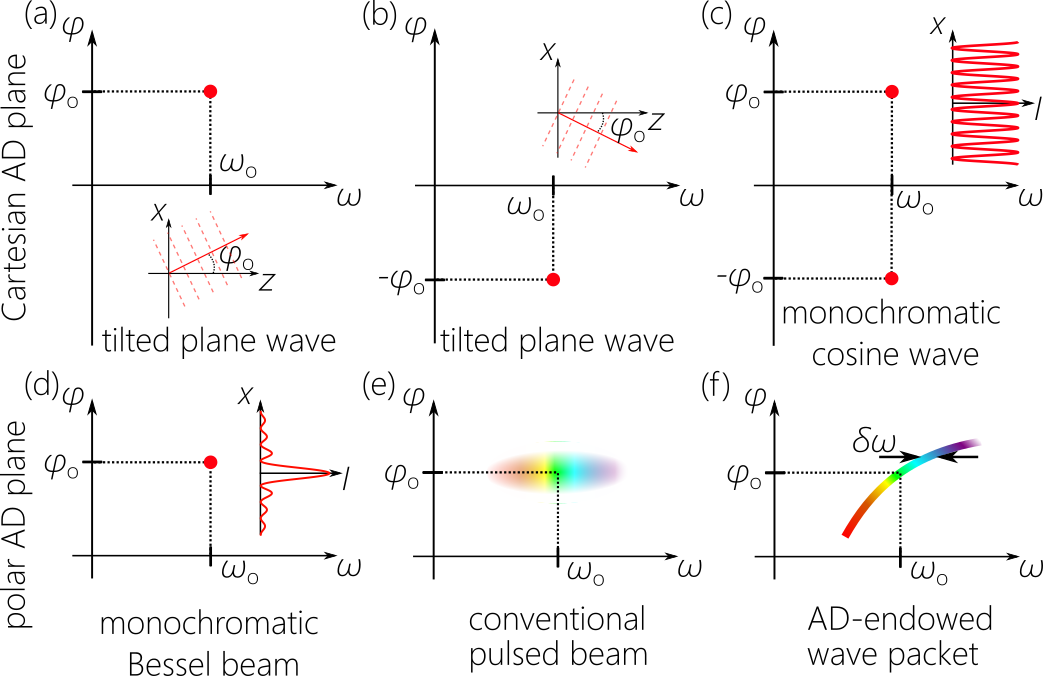}
\caption{(a) A single point $(\omega_{\mathrm{o}},\varphi_{\mathrm{o}})$ in the upper half of the \textit{Cartesian} AD plane represents a monochromatic plane wave $\exp\{ik_{\mathrm{o}}(x\sin\varphi_{\mathrm{o}}+z\cos\varphi_{\mathrm{o}}-ct)\}$, depicted as an inset. (b) A single point $(\omega_{\mathrm{o}},-\varphi_{\mathrm{o}})$ in the lower half of the Cartesian AD plane represents a monochromatic plane wave $\exp\{ik_{\mathrm{o}}(-x\sin\varphi_{\mathrm{o}}+z\cos\varphi_{\mathrm{o}}-ct)\}$, depicted as an inset. (c) A pair of points $(\omega_{\mathrm{o}},\varphi_{\mathrm{o}})$ and $(\omega_{\mathrm{o}},-\varphi_{\mathrm{o}})$ in the Cartesian AD plane identifies a monochromatic cosine beam $\cos\left\{k_{\mathrm{o}}x\sin\varphi_{\mathrm{o}}\right\}\exp\{ik_{\mathrm{o}}(z\cos\varphi_{\mathrm{o}}-ct)\}$, depicted as an inset. (d) A single point $(\omega_{\mathrm{o}},\varphi_{\mathrm{o}})$ in the \textit{polar} (circularly symmetric) AD plane represents a monochromatic Bessel beam $J_{0}(k_{\mathrm{o}}r\sin\varphi_{\mathrm{o}})\exp\{ik_{\mathrm{o}}(z\cos\varphi_{\mathrm{o}}-ct)\}$, depicted as an inset. (e) A conventional pulsed beam (or wave packet) corresponds to a finite 2D domain in the AD plane (whether Cartesian or polar). (f) An AD-endowed field corresponds to a 1D trajectory in the AD plane (whether Cartesian or polar). The inevitable spectral uncertainty $\delta\omega$ gives a finite thickness to this curve; that is, each angle $\varphi$ is associated with a finite spectral bandwidth $\delta\omega$.}
\label{fig:Definition}
\end{figure}

We will be making use throughout of the depiction of optical fields in an AD plane that is defined by two axes: the frequency $\omega$ and the propagation angle $\varphi$ [Fig.~\ref{fig:Definition}]. We distinguish between two different AD planes: the Cartesian AD plane and the polar AD plane. The Cartesian AD plane [Fig.~\ref{fig:Definition}(a-c)] is suitable for fields in which AD is introduced along one transverse dimension (taken here to be $x$). This is the typical field configuration produced by diffraction gratings and prisms. The propagation angle can take on both positive and negative values, corresponding to propagation on opposite sides of the $z$-axis along~$x$. We take the field to be either uniform along $y$ or at least that the field is separable with respect to $x$ and $y$. A point $(\omega_{\mathrm{o}},\varphi_{\mathrm{o}})$ in the upper half of the Cartesian AD plane identifies a monochromatic plane wave of the form $e^{ik_{\mathrm{o}}(x\sin\varphi_{\mathrm{o}}+z\cos\varphi_{\mathrm{o}}-ct)}$ [Fig.~\ref{fig:Definition}(a)], whereas a point $(\omega_{\mathrm{o}},-\varphi_{\mathrm{o}})$ in the lower half of the Cartesian AD plane identifies a monochromatic plane wave of the form $e^{ik_{\mathrm{o}}(-x\sin\varphi_{\mathrm{o}}+z\cos\varphi_{\mathrm{o}}-ct)}$ [Fig.~\ref{fig:Definition}(b)]. Consequently, symmetrically placed points $(\omega_{\mathrm{o}},\varphi_{\mathrm{o}})$ and $(\omega_{\mathrm{o}},-\varphi_{\mathrm{o}})$ in the Cartesian AD plane correspond to a monochromatic cosine wave $\cos\left(k_{\mathrm{o}}x\sin\varphi_{\mathrm{o}}\right)e^{ik_{\mathrm{o}}(z\cos\varphi_{\mathrm{o}}-ct)}$ [Fig.~\ref{fig:Definition}(c)].

The polar AD plane is suitable for circularly symmetric fields; i.e., the transverse field distribution in any axial plane in polar coordinates $(r,\phi)$ depends solely on the radial coordinate $r$, in which case the propagation angles $\varphi$ can take on only positive values. A single point $(\omega_{\mathrm{o}},\varphi_{\mathrm{o}})$ in this polar AD plane therefore represents a monochromatic Bessel beam $J_{0}(k_{\mathrm{o}}r\sin\varphi_{\mathrm{o}})e^{ik_{\mathrm{o}}(z\cos\varphi_{\mathrm{o}}-ct)}$, corresponding to the frequency $\omega_{\mathrm{o}}$ propagating in a conical configuration with apex angle $\varphi_{\mathrm{o}}$ [Fig.~\ref{fig:Definition}(d)], where $J_{0}(\cdot)$ is the $0^{\mathrm{th}}$-order Bessel function of the first kind.

The representation of the optical field in the AD plane is quite general, with $-90^{\circ}<\varphi<90^{\circ}$ in the Cartesian AD plane, and $0<\varphi<90^{\circ}$ in the (circularly symmetric) polar AD plane. Within this conception, a conventional pulsed beam is represented by a finite 2D domain in the AD plane (whether Cartesian or polar) [Fig.~\ref{fig:Definition}(e)]. In this paper we are concerned with optical fields endowed with AD. Because each frequency $\omega$ propagates at a different angle $\varphi(\omega)$ in this scenario, the angular spectrum of an AD-endowed field is represented in the AD plane by a 1D curve [Fig.~\ref{fig:Definition}(f)]. Of course, a strictly defined 1D curve in the AD plane corresponds to a field having infinite energy. In practice, there is always a finite `spectral uncertainty' $\delta\omega$ in the relationship between $\omega$ and $\varphi$, which typically arises from finite apertures. In other words, there is always some `fuzziness' in the curve in Fig.~\ref{fig:Definition}(f) that gives it a finite `thickness' \cite{Yessenov19OE,Kondakci19OL}.

\begin{figure}[t!]
\centering
\includegraphics[width=8.6cm]{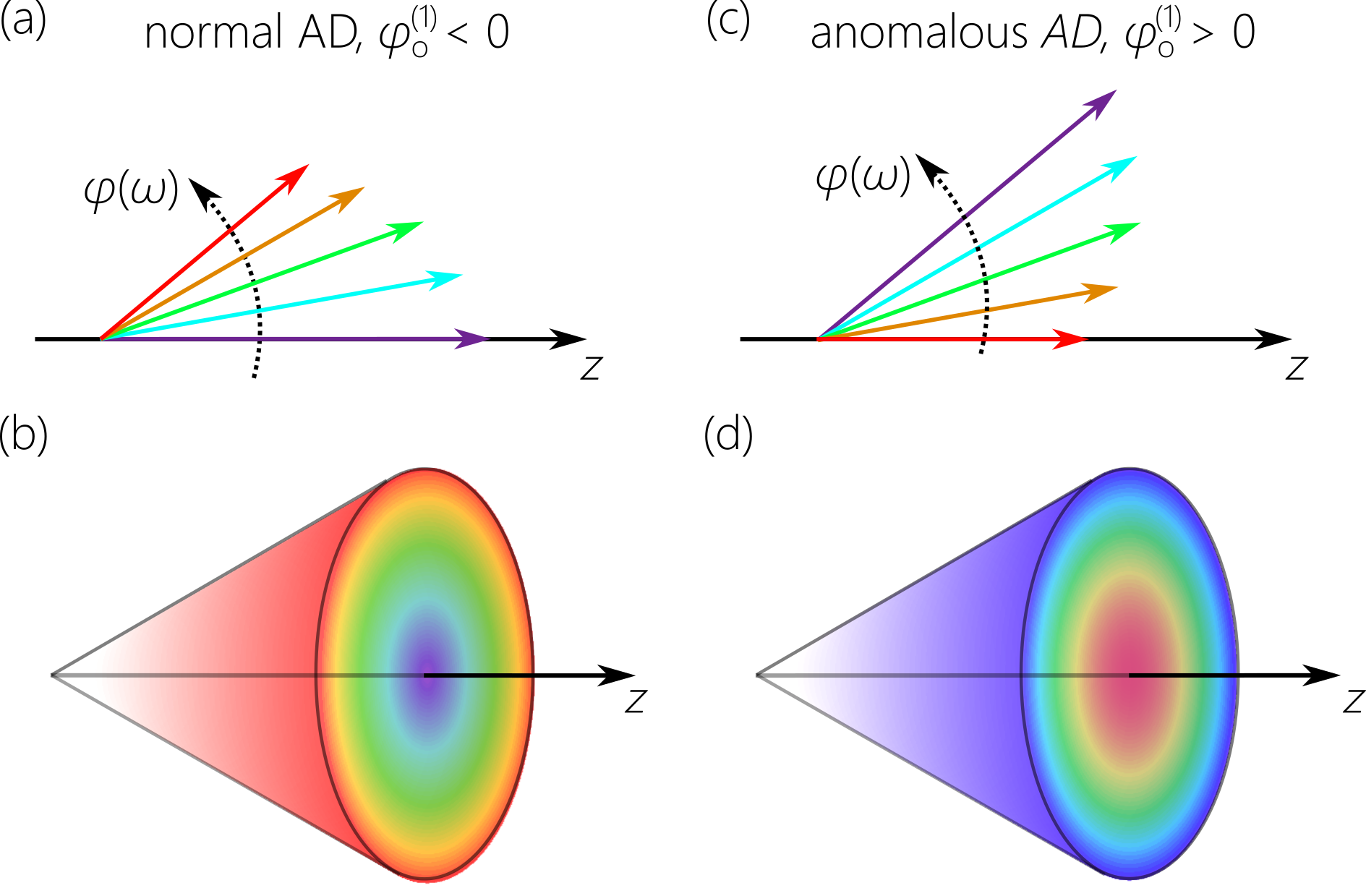}
\caption{(a) Normal AD $\varphi_{\mathrm{o}}^{(1)}<0$ is associated with lower frequencies (longer wavelengths; red) having larger propagation angles, while higher frequencies (shorter wavelengths; blue) travel at smaller angles. (b) Azimuthally symmetric, conical spectral depiction of (a). (c) Anomalous AD $\varphi_{\mathrm{o}}^{(1)}>0$ is associated with higher frequencies (shorter wavelengths; blue) having larger propagation angles, while lower frequencies (longer wavelengths; red) travel at smaller angles. (d) Azimuthally symmetric, conical spectral depiction of (c).}
\label{fig:FlippingTheADSign}
\end{figure}

\subsection{Normal and anomalous AD}

It is useful to define two regimes for the AD coefficient $\varphi_{\mathrm{o}}^{(1)}$. We refer to $\varphi_{\mathrm{o}}^{(1)}<0$ as `normal AD' [Fig.~\ref{fig:FlippingTheADSign}(a)] and to $\varphi_{\mathrm{o}}^{(1)}>0$ as `anomalous AD' [Fig.~\ref{fig:FlippingTheADSign}(b)] -- also referred to as reverse-color-mode sequence \cite{Bartl4PNAS}. There are three reasons that motivate this nomenclature. First, if broadband light is incident normally on a diffraction grating, longer wavelengths diffract at larger angles with respect to the grating normal, while shorter wavelengths diffract at smaller angles with the normal [Fig.~\ref{fig:FlippingTheADSign}(a,b)]. When the reverse-color-mode sequence in Fig.~\ref{fig:FlippingTheADSign}(c,d) occurs in the context of a grating, this indicates deviation from normal incidence. Such a configuration has been well-documented in nature; e.g., in diffraction from the \textit{Pierella luna} butterfly wing \cite{Vigneron10PRE,England14PNAS,Bartl4PNAS}. The same effect has been exploited in resonant photonics to broaden  the coupling bandwidth to Fabry-P{\'e}rot resonators (so-called `omni-resonance' \cite{Shabahang17SR,Shiri20OL,Hall25LRR}), whose resonances are associated with anomalous AD. The second motivation for the nomenclature in Fig.~\ref{fig:FlippingTheADSign} stems from the spectral changes that are incurred by a freely propagating polychromatic optical field upon diffraction. The longer wavelengths diffract faster and are observed downstream at the field outer edge. Once again, the form of the field in Fig.~\ref{fig:FlippingTheADSign}(c,d) does not occur normally in the course of diffraction. Thirdly, this nomenclature matches the conventional definition for chromatic dispersion where `normal dispersion' refers to a refractive index that increases with frequency and `anomalous dispersion' to the opposite trend (see Part~II of this tutorial \cite{Hall26JOSAA2}).

\subsection{Symmetrized AD-endowed fields}

\begin{figure}[t!]
\centering
\includegraphics[width=8.6cm]{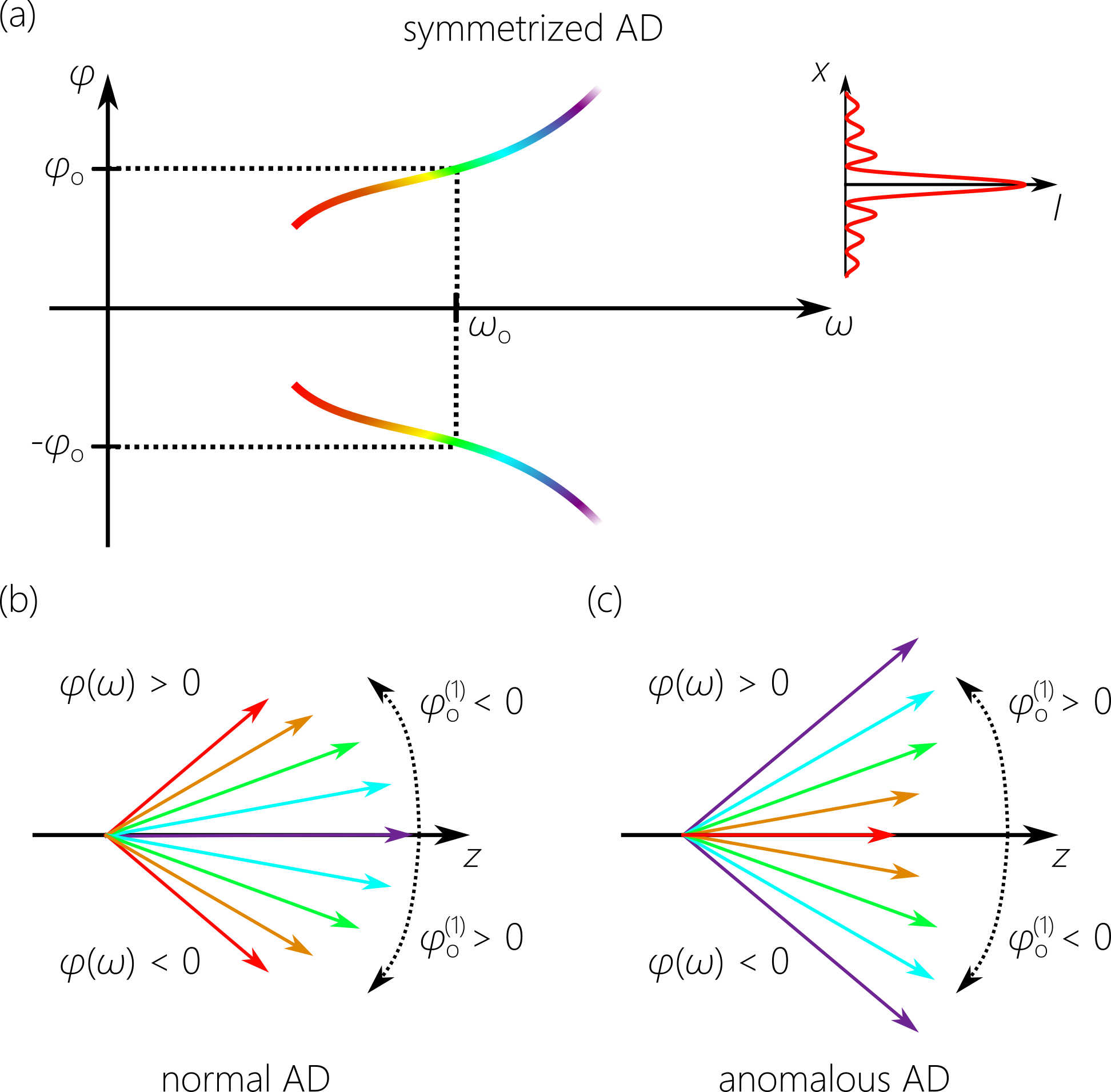}
\caption{(a) A symmetrized AD profile in the Cartesian AD plane. The AD spectrum has mirror symmetry around the $\omega$-axis, and the time-averaged spatial profile $I(x)$ has an on-axis peak (inset). (b) Normal AD and (c) anomalous AD for symmetrized fields in the Cartesian AD plane.}
\label{fig:Symmetrized}
\end{figure}

We consider here symmetrized AD spectra that yield an on-axis peak, which is monitored along the propagation axis (typically with respect to a reference pulse) to estimate $\widetilde{v}$. This occurs naturally in the polar AD plane [Fig.~\ref{fig:Definition}(d)] where each point in the AD plane corresponds to a Bessel beam with an on-axis peak. In the Cartesian AD plane we require that the AD profile have mirror symmetry around the $\omega$-axis [Fig.~\ref{fig:Symmetrized}(a)]. In other words, each frequency $\omega$ is associated with two propagation angles $\varphi(\omega)$ and $-\varphi(\omega)$, which constitute a monochromatic cosine wave. The field is thus a superposition of such cosine waves that together yield an on-axis peak [Fig.~\ref{fig:Symmetrized}(a), inset].

The definition of normal AD, $\varphi_{\mathrm{o}}^{(1)}<0$, is valid for $\varphi>0$ in both the polar and Cartesian AD planes. When $\varphi<0$, the mirror symmetry around the $\omega$-axis indicates that normal AD is associated with $\varphi_{\mathrm{o}}^{(1)}>0$ [Fig.~\ref{fig:Symmetrized}(b)], and vice versa for anomalous AD [Fig.~\ref{fig:Symmetrized}(c)]: $\varphi_{\mathrm{o}}^{(1)}>0$ for $\varphi>0$, and $\varphi_{\mathrm{o}}^{(1)}<0$ for $\varphi<0$.

\section{Group velocity in presence of angular dispersion}

\subsection{Definition of group velocity in terms of AD}

\begin{figure}[t!]
\centering
\includegraphics[width=8.6cm]{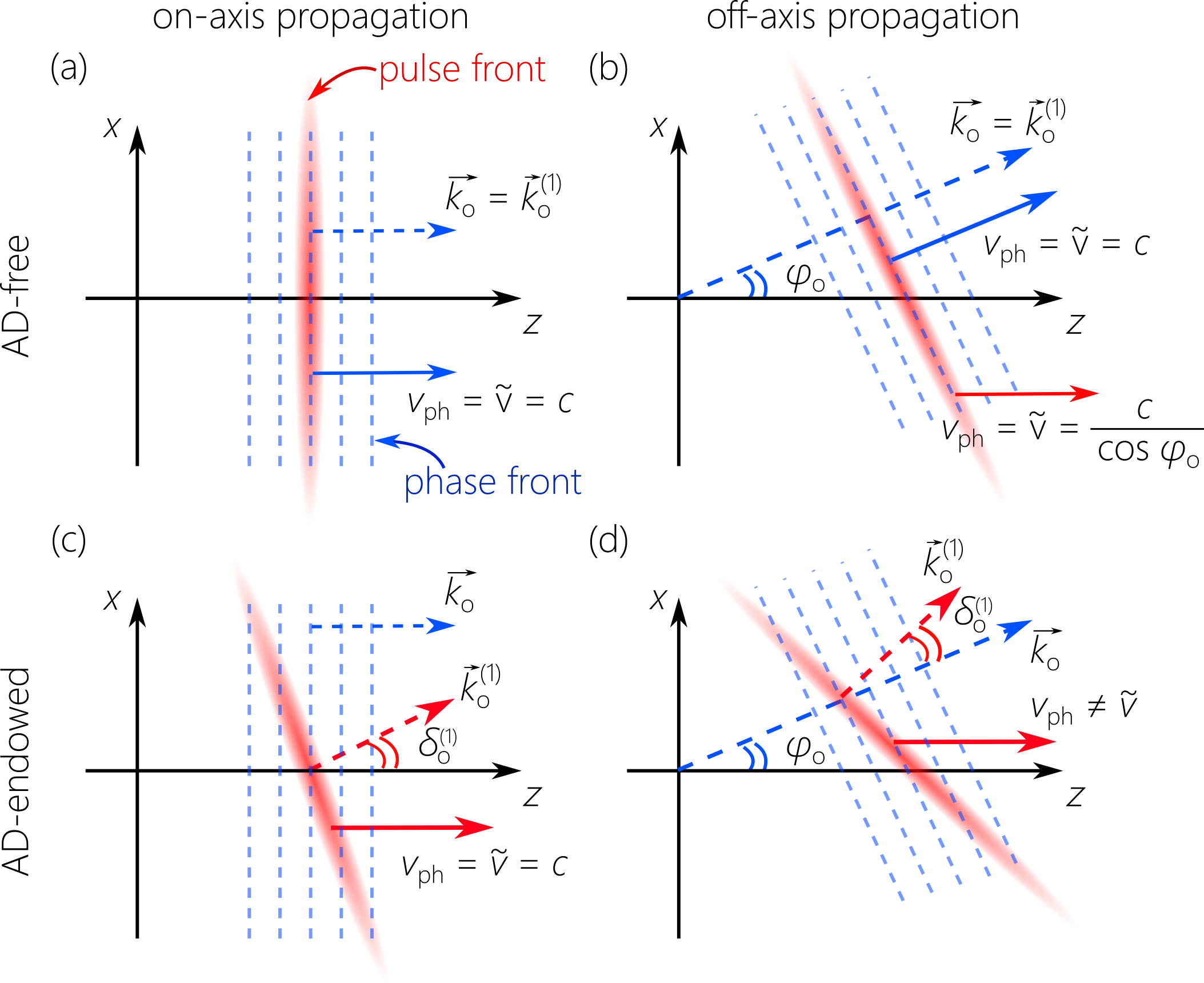}
\caption{Phase and group velocities in presence of AD. (a) For an on-axis collimated field ($\varphi_{\mathrm{o}}=0$ and $\varphi_{\mathrm{o}}^{(1)}=0$), $v_{\mathrm{ph}}=\widetilde{v}=c$ along the $z$-axis. (b) For an off-axis collimated field ($\varphi_{\mathrm{o}}\neq0$ and $\varphi_{\mathrm{o}}^{(1)}=0$), $v_{\mathrm{ph}}=\widetilde{v}=c/\cos\varphi_{\mathrm{o}}>c$ along the $z$-axis. This is a purely geometric effect. Along $\vec{k}_{\mathrm{o}}$ (the normal to the phase front), $v_{\mathrm{ph}}=\widetilde{v}=c$. In absence of AD, the pulse front and the phase front coincide. (c) For an on-axis field endowed with AD ($\varphi_{\mathrm{o}}=0$ and $\varphi_{\mathrm{o}}^{(1)}\neq0$), the pulse front is tilted an angle $\delta_{\mathrm{o}}^{(1)}$ with the phase front, where $\tan\delta_{\mathrm{o}}^{(1)}=\omega_{\mathrm{o}}\varphi_{\mathrm{o}}^{(1)}$. Along the $z$-axis, $v_{\mathrm{ph}}=\widetilde{v}=c$. (d) For an off-axis field in presence of AD ($\varphi_{\mathrm{o}}\neq0$ and $\varphi_{\mathrm{o}}^{(1)}\neq0$), $\widetilde{v}$ along the $z$-axis is given by Eq.~\ref{eq:GroupVelocityGeneral}, and $\widetilde{v}=c$ along $\vec{k}_{\mathrm{o}}$. In (c,d), $\widetilde{v}$ along $\vec{k}_{\mathrm{o}}^{(1)}$ is given by Eq.~\ref{eq:GroupVelAlongko1}.}
\label{fig:PhaseAndGroup}
\end{figure}

The \textit{phase velocity} $v_{\mathrm{ph}}$ along the $z$-axis is simply:
\begin{equation}
v_{\mathrm{ph}}=\frac{\omega}{k_{z}}\biggr|_{\omega_{\mathrm{o}}}=\frac{c}{\cos\varphi_{\mathrm{o}}}.
\end{equation}
Therefore, on-axis fields ($\varphi_{\mathrm{o}}=0$) have $v_{\mathrm{ph}}=c$ along the $z$-axis whether AD is absent or present [Fig.~\ref{fig:PhaseAndGroup}(a,c)]. Any deviation in $v_{\mathrm{ph}}$ along the $z$-axis from $c$ is thus a purely \textit{geometric} effect as a consequence of the tilted propagation angle $\varphi_{\mathrm{o}}$ [Fig.~\ref{fig:PhaseAndGroup}(b,d)].

The \textit{group velocity} $\widetilde{v}$ along the $z$-axis, $\tfrac{1}{\widetilde{v}}\!=\!\tfrac{dk_{z}}{d\omega}\big|_{\omega_{\mathrm{o}}}$, is \cite{Porras03PRE2,Hall25APLP}:
\begin{equation}\label{eq:GroupVelocityGeneral}
\widetilde{v}=\frac{c}{\cos\varphi_{\mathrm{o}}-\omega_{\mathrm{o}}\varphi_{\mathrm{o}}^{(1)}\sin\varphi_{\mathrm{o}}}=\frac{v_{\mathrm{ph}}}{1-\omega_{\mathrm{o}}\varphi_{\mathrm{o}}^{(1)}\tan\varphi_{\mathrm{o}}},
\end{equation}
which depends solely on $\varphi_{\mathrm{o}}$ and $\varphi_{\mathrm{o}}^{(1)}$. The higher-order AD expansion coefficients in Eq.~\ref{eq:ADexpansion} do \textit{not} contribute to $\widetilde{v}$. It is customary to denote the first-order AD in terms of an angle: $\omega_{\mathrm{o}}\varphi_{\mathrm{o}}^{(1)}\!=\!\tan\delta_{\mathrm{o}}^{(1)}$, where $\delta_{\mathrm{o}}^{(1)}$ is the tilt angle between the vector $\vec{k}_{\mathrm{o}}$ that is normal to the phase front (the plane of constant phase) and the vector $\vec{k}_{\mathrm{o}}^{(1)}$ that is normal to the pulse front (the plane of constant intensity) \cite{Torres10AOP,Fulop10Review,Hall21OL,Hall22OEConsequences}; see Fig.~\ref{fig:PhaseAndGroup}(c,d). The group velocity can then be expressed solely in terms of $\varphi_{\mathrm{o}}$ and $\delta_{\mathrm{o}}^{(1)}$:
\begin{equation}
\widetilde{v}=c\frac{\cos\delta_{\mathrm{o}}^{(1)}}{\cos\left(\varphi_{\mathrm{o}}+\delta_{\mathrm{o}}^{(1)}\right)}.
\end{equation}
In absence of AD [Fig.~\ref{fig:PhaseAndGroup}(a,b)], the pulse and phase fronts coincide, and the vectors $\vec{k}_{\mathrm{o}}$ and $\vec{k}_{\mathrm{o}}^{(1)}$ thus also coincide.

The formula in Eq.~\ref{eq:GroupVelocityGeneral} is central to us here and deserves to be better known. It indicates that the deviation in $\widetilde{v}$ along the $z$-axis from $c$ in free space is due to the combined impact of two factors:
\begin{enumerate}
\item a \textit{geometric} factor that stems from the tilted propagation angle $\varphi_{\mathrm{o}}$ with respect to the $z$-axis; and 
\item an \textit{interferometric} effect resulting from the first-order AD term $\varphi_{\mathrm{o}}^{(1)}$.
\end{enumerate}
This gives rise to four distinct field configurations:
\begin{enumerate}
\item On-axis AD-free, $\varphi_{\mathrm{o}}=0$ and $\varphi_{\mathrm{o}}^{(1)}=0$ [Fig.~\ref{fig:PhaseAndGroup}(a)].
\item Off-axis AD-free, $\varphi_{\mathrm{o}}\neq0$ and $\varphi_{\mathrm{o}}^{(1)}=0$ [Fig.~\ref{fig:PhaseAndGroup}(b)].
\item On-axis AD-endowed, $\varphi_{\mathrm{o}}=0$ and $\varphi_{\mathrm{o}}^{(1)}\neq0$ [Fig.~\ref{fig:PhaseAndGroup}(c)].
\item Off-axis AD-endowed, $\varphi_{\mathrm{o}}\neq0$ and $\varphi_{\mathrm{o}}^{(1)}\neq0$ [Fig.~\ref{fig:PhaseAndGroup}(d)].
\end{enumerate}

Off-axis propagation is a prerequisite for tuning $\widetilde{v}$ [Fig.~\ref{fig:PhaseAndGroup}(b,d)]. Therefore, for on-axis fields ($\varphi_{\mathrm{o}}=0$), we strictly have $\widetilde{v}=c$ along the $z$-axis (in which case it coincides with the propagation axis), and $\varphi_{\mathrm{o}}^{(1)}$ has no effect on $\widetilde{v}$ [Fig.~\ref{fig:PhaseAndGroup}(a,c)]. In absence of AD, $\widetilde{v}=v_{\mathrm{ph}}=c/\cos\varphi_{\mathrm{o}}$, and deviation from $c$ is purely a geometric effect [Fig.~\ref{fig:PhaseAndGroup}(b)]. In presence of both off-axis propagation \textit{and} AD [Fig.~\ref{fig:PhaseAndGroup}(d)], the confluence of these effects can yield arbitrary group velocities in free space.

Therefore, in Fig.~\ref{fig:PhaseAndGroup}(a) where $\varphi_{\mathrm{o}}=0$ and $\varphi_{\mathrm{o}}^{(1)}=0$ (on-axis, AD-free field), the phase and pulse fronts coincide, and $\vec{k}_{\mathrm{o}}$ and $\vec{k}_{\mathrm{o}}^{(1)}$ in turn coincide with each other and with the $z$-axis. This implies that $v_{\mathrm{ph}}=\widetilde{v}=c$ along the $z$-axis. In Fig.~\ref{fig:PhaseAndGroup}(b) where $\varphi_{\mathrm{o}}\neq0$ but $\varphi_{\mathrm{o}}^{(1)}=0$ (off-axis, AD-free field), the phase and pulse fronts coincide, $\vec{k}_{\mathrm{o}}$ and $\vec{k}_{\mathrm{o}}^{(1)}$ in turn coincide with each other, and both make an angle $\varphi_{\mathrm{o}}$ with the $z$-axis. Therefore, $v_{\mathrm{ph}}=\widetilde{v}=c/\cos\varphi_{\mathrm{o}}$ along the $z$-axis, while $v_{\mathrm{ph}}=\widetilde{v}=c$ along $\vec{k}_{\mathrm{o}}$ and $\vec{k}_{\mathrm{o}}^{(1)}$. The group velocity can be tuned via the geometric factor $\cos\varphi_{\mathrm{o}}$, but $\widetilde{v}$ does not deviate substantially from $c$ except in the non-paraxial regime (large $\varphi_{\mathrm{o}}$). In Fig.~\ref{fig:PhaseAndGroup}(c) where $\varphi_{\mathrm{o}}=0$ but $\varphi_{\mathrm{o}}^{(1)}\neq0$ (on-axis AD-endowed field), the phase front is tilted an angle $\delta_{\mathrm{o}}^{(1)}$ to the pulse front, $\vec{k}_{\mathrm{o}}$ coincides with the $z$-axis, and $\vec{k}_{\mathrm{o}}^{(1)}$ makes an angle $\delta_{\mathrm{o}}^{(1)}$ with $\vec{k}_{\mathrm{o}}$. Therefore $v_{\mathrm{ph}}=\widetilde{v}=c$ along the $z$-axis (which coincides with $\vec{k}_{\mathrm{o}}$), but along $\vec{k}_{\mathrm{o}}^{(1)}$ we have $v_{\mathrm{ph}}=c/\cos\delta_{\mathrm{o}}^{(1)}$ and:
\begin{equation}\label{eq:GroupVelAlongko1}
\widetilde{v}=\frac{c}{\sqrt{1+(\omega_{\mathrm{o}}\varphi_{\mathrm{o}}^{(1)})^{2}}}\leq c.
\end{equation}
The wave packet in this configuration is known as a tilted pulse front (TPF) \cite{Torres10AOP,Fulop10Review,Hall22OEConsequences}.

Finally, in Fig.~\ref{fig:PhaseAndGroup}(d) where $\varphi_{\mathrm{o}}\neq0$ and $\varphi_{\mathrm{o}}^{(1)}\neq0$ (off-axis, AD-endowed field), the phase and pulse fronts are tilted an angle $\delta_{\mathrm{o}}^{(1)}$ with respect to each other, $\vec{k}_{\mathrm{o}}$ makes an angle $\varphi_{\mathrm{o}}$ with the $z$-axis, and $\vec{k}_{\mathrm{o}}^{(1)}$ makes an angle $\varphi_{\mathrm{o}}+
\delta_{\mathrm{o}}^{(1)}$ with the $z$-axis. Therefore, $v_{\mathrm{ph}}=c/\cos\varphi_{\mathrm{o}}$ along the $z$ axis, $v_{\mathrm{ph}}=c$ along $\vec{k}_{\mathrm{o}}$, and $v_{\mathrm{ph}}=c/\cos\delta_{\mathrm{o}}^{(1)}$ along $\vec{k}_{\mathrm{o}}^{(1)}$; whereas $\widetilde{v}$ along the $z$-axis is given by Eq.~\ref{eq:GroupVelocityGeneral}, $\widetilde{v}=c$ along $\vec{k}_{\mathrm{o}}$, and $\widetilde{v}$ along $\vec{k}_{\mathrm{o}}^{(1)}$ is given in Eq.~\ref{eq:GroupVelAlongko1}. Note that $\widetilde{v}$ is always luminal ($\widetilde{v}=c$) along $\vec{k}_{\mathrm{o}}$, and is always subluminal ($\widetilde{v}\leq c$) along $\vec{k}_{\mathrm{o}}^{(1)}$.

Crucially, the only assumption underpinning the result in Eq.~\ref{eq:GroupVelocityGeneral} is the validity of the AD expansion in Eq.~\ref{eq:ADexpansion}, which necessitates that $\varphi(\omega)$ be differentiable at $\omega_{\mathrm{o}}$; i.e., $\varphi_{\mathrm{o}}^{(1)}$ is well-defined. We thus refer to conventional forms of AD as `differentiable AD'. We discuss in Part~II of this tutorial \cite{Hall26JOSAA2} the consequences of an AD spectral profile where $\varphi_{\mathrm{o}}^{(1)}$ is \textit{not} defined, which we refer to as `non-differentiable AD'.

The need to have off-axis propagation ($\varphi_{\mathrm{o}}\neq0$) in order to tune $\widetilde{v}$ can be a severe restriction on the distance over which the group velocity is maintained. One may in principle reduce $\varphi_{\mathrm{o}}$ to monitor $\widetilde{v}$ over a larger distance while maintaining its value by increasing $\varphi_{\mathrm{o}}^{(1)}$ according to Eq.~\ref{eq:GroupVelocityGeneral}. However, this is not always possible. Indeed, in most conventional optical settings, $\varphi_{\mathrm{o}}$ and $\varphi_{\mathrm{o}}^{(1)}$ are \textit{not} physically independent of each other. Furthermore, the value of $\omega_{\mathrm{o}}\varphi_{\mathrm{o}}^{(1)}$ required to accommodate small values of $\varphi_{\mathrm{o}}$ is usually beyond the capabilities of conventional optical components \cite{Hall25APLP}.

\subsection{Different regimes for the group velocity}

\begin{figure}[t!]
\centering
\includegraphics[width=8.6cm]{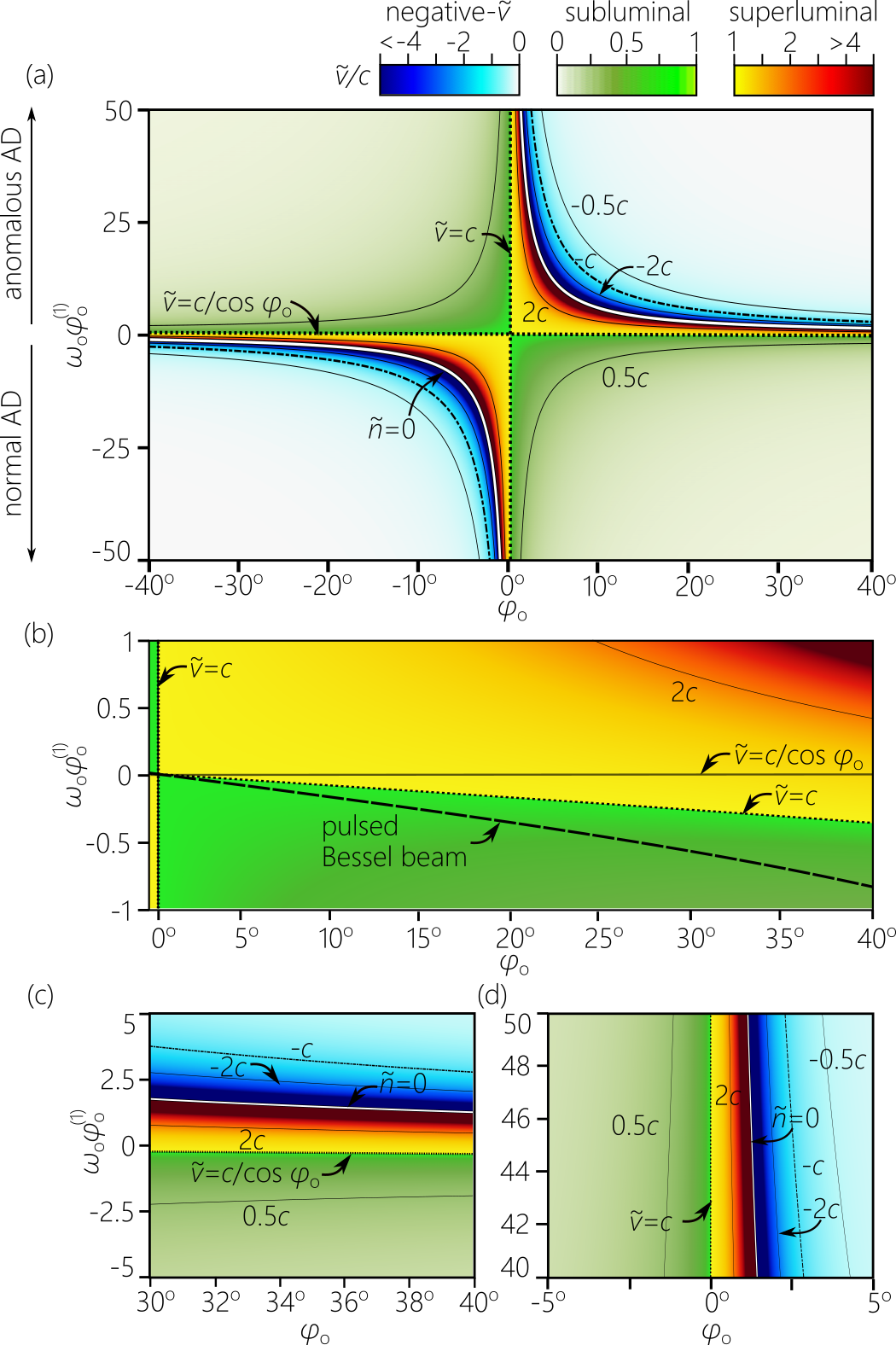}
\caption{(a) The group velocity $\widetilde{v}$ of a pulsed beam endowed with AD as a function of $\varphi_{\mathrm{o}}$ and $\omega_{\mathrm{o}}\varphi_{\mathrm{o}}^{(1)}$. We use different color palettes to distinguish the superluminal, subluminal, and negative-$\widetilde{v}$ regimes. The black curves are contours of fixed $\widetilde{v}$, and the white curve identifies the condition $\widetilde{n}=0$. (b-d) Expanded views of sections of (a). In panel (b), we identify two contours. The first contour is the luminal contour $\widetilde{n}=1$ ($\widetilde{v}=c$) given in Eq.~\ref{eq:LuminalContour}, which corresponds to the condition $\omega_{\mathrm{o}}\varphi_{\mathrm{o}}^{(1)}=-\tan\tfrac{\varphi_{\mathrm{o}}}{2}$. This contour bounds a superluminal area that extends within the normal AD regime. The second contour corresponds to the condition $\omega_{\mathrm{o}}\varphi_{\mathrm{o}}^{(1)}=-\tan\varphi_{\mathrm{o}}$, which identifies subluminal pulsed Bessel beams \cite{Giovannini15Science}.}
\label{fig:GroupVelocity}
\end{figure}

We plot in Fig.~\ref{fig:GroupVelocity}(a) the group velocity $\widetilde{v}$ in terms of $\varphi_{\mathrm{o}}$ and $\omega_{\mathrm{o}}\varphi_{\mathrm{o}}^{(1)}$ using Eq.~\ref{eq:GroupVelocityGeneral}; in the polar AD plane, only the right half of the plot ($\varphi_{\mathrm{o}}>0$) is relevant. Note that $\widetilde{v}=c$ along the vertical axis at $\varphi_{\mathrm{o}}=0$, and $\widetilde{v}=c/\cos\varphi_{\mathrm{o}}$ along the horizontal axis at $\omega_{\mathrm{o}}\varphi_{\mathrm{o}}^{(1)}=0$ [Fig.~\ref{fig:GroupVelocity}(a)]. The upper right quadrant corresponds to superluminal and negative-$\widetilde{v}$ regimes, whereas the lower right quadrant corresponds mainly to the subluminal regime. The plot has inversion symmetry around the origin; i.e., the points $(\varphi_{\mathrm{o}},\omega_{\mathrm{o}}\varphi_{\mathrm{o}}^{(1)})$ and $(-\varphi_{\mathrm{o}},-\omega_{\mathrm{o}}\varphi_{\mathrm{o}}^{(1)})$ yield the same $\widetilde{v}$.

It is clear that $\widetilde{v}$ can, in principle, take on arbitrary values if one has \textit{independent} control over $\varphi_{\mathrm{o}}$ and $\omega_{\mathrm{o}}\varphi_{\mathrm{o}}^{(1)}$, \textit{and} one can produce arbitrary values of $\omega_{\mathrm{o}}\varphi_{\mathrm{o}}^{(1)}$. To identify the different regimes for $\widetilde{v}$ in free space, we define the group index:
\begin{equation}\label{eq:GeneralGroupIndex}
\widetilde{n}=c/\widetilde{v}=\cos\varphi_{\mathrm{o}}-\omega_{\mathrm{o}}\varphi_{\mathrm{o}}^{(1)}\sin\varphi_{\mathrm{o}};
\end{equation}
$\widetilde{n}=1$ is luminal, $\widetilde{n}>1$ is subluminal, $0<\widetilde{n}<1$ is superluminal, and $\widetilde{n}<0$ corresponds to negative group velocity. Re-writing this equation as $\omega_{\mathrm{o}}\varphi_{\mathrm{o}}^{(1)}=\tfrac{\cos\varphi_{\mathrm{o}}-\widetilde{n}}{\sin\varphi_{\mathrm{o}}}$ entails that the contours of fixed $\widetilde{n}$ have approximately the form $\omega_{\mathrm{o}}\varphi_{\mathrm{o}}^{(1)}\sim\tfrac{1}{\varphi_{\mathrm{o}}}$ at small angles. In other words, realizing a particular $\widetilde{v}$ at a small angle $\varphi_{\mathrm{o}}$ requires a large AD value $\omega_{\mathrm{o}}\varphi_{\mathrm{o}}^{(1)}$ [Fig.~\ref{fig:GroupVelocity}(b)], while operating at large $\varphi_{\mathrm{o}}$ reduces the AD value required [Fig.~\ref{fig:GroupVelocity}(c)]. 

We start by examining a few special cases for $\varphi_{\mathrm{o}}\geq0$ as signposts. First, at $\varphi_{\mathrm{o}}=0$, we have $\widetilde{n}=1$ and the wave packet is on-axis and luminal independently of the value of $\omega_{\mathrm{o}}\varphi_{\mathrm{o}}^{(1)}$. Second, at $\varphi_{\mathrm{o}}\rightarrow90^{\circ}$ we have $\widetilde{n}\rightarrow-\omega_{\mathrm{o}}\varphi_{\mathrm{o}}^{(1)}$, so that the wave packet is subluminal in the normal-AD regime ($\omega_{\mathrm{o}}\varphi_{\mathrm{o}}^{(1)}<0$) when $|\omega_{\mathrm{o}}\varphi_{\mathrm{o}}^{(1)}|>1$, is superluminal in the normal-AD regime when $|\omega_{\mathrm{o}}\varphi_{\mathrm{o}}^{(1)}|<1$, and has a negative-$\widetilde{v}$ in the anomalous-AD regime ($\omega_{\mathrm{o}}\varphi_{\mathrm{o}}^{(1)}>0$). Third, for $\omega_{\mathrm{o}}\varphi_{\mathrm{o}}^{(1)}=0$ we have $\widetilde{n}=\cos\varphi_{\mathrm{o}}$, and the AD-free wave packet is superluminal (corresponding to the X-waves described below). Except for large values of $\varphi_{\mathrm{o}}$, $\widetilde{v}\approx c$ along the horizontal axis $\omega_{\mathrm{o}}\varphi_{\mathrm{o}}^{(1)}=0$.

\textit{Luminal regime, $\widetilde{v}=c$ ($\widetilde{n}=1$).} The luminal condition $\widetilde{v}=c$ occurs in two different scenarios: (1) on-axis propagation ($\varphi_{\mathrm{o}}=0$) independently of the value of $\varphi_{\mathrm{o}}^{(1)}$, corresponding to the vertical axis in Fig.~\ref{fig:GroupVelocity}(a); and (2) off-axis propagation when
\begin{equation}\label{eq:LuminalContour}
\omega_{\mathrm{o}}\varphi_{\mathrm{o}}^{(1)}=\frac{\cos\varphi_{\mathrm{o}}-1}{\sin\varphi_{\mathrm{o}}}=-\tan\frac{\varphi_{\mathrm{o}}}{2},
\end{equation}
which requires independent control over $\varphi_{\mathrm{o}}$ and $\varphi_{\mathrm{o}}^{(1)}$. In general, $\varphi_{\mathrm{o}}^{(1)}<0$ is required when $\varphi_{\mathrm{o}}>0$ to achieve $\widetilde{v}=c$, while $\varphi_{\mathrm{o}}^{(1)}>0$ is required when $\varphi_{\mathrm{o}}<0$. For $\varphi_{\mathrm{o}}\geq0$, this curve is located in the bottom right quadrant, as depicted in the expanded view in Fig.~\ref{fig:GroupVelocity}(b).

\textit{Subluminal regime, $\widetilde{v}<c$ $(\widetilde{n}>1)$.} Taking $\varphi_{\mathrm{o}}>0$, normal AD ($\omega_{\mathrm{o}}\varphi_{\mathrm{o}}^{(1)}<0$) is required to produce a subluminal group velocity. This constitutes most of the lower right quadrant of Fig.~\ref{fig:GroupVelocity}(a), which is bounded from above by the luminal condition $\omega_{\mathrm{o}}\varphi_{\mathrm{o}}^{(1)}=-\tan\tfrac{\varphi_{\mathrm{o}}}{2}$ [Fig.~\ref{fig:GroupVelocity}(b)]. 

\textit{Superluminal regime, $\widetilde{v}>c$ $(\widetilde{n}<1)$.} The superluminal regime extends between the luminal $\widetilde{n}=1$ contour corresponding to $\omega_{\mathrm{o}}\varphi_{\mathrm{o}}^{(1)}=-\tan\tfrac{\varphi_{\mathrm{o}}}{2}$ [Fig.~\ref{fig:GroupVelocity}(b)] and the contour $\widetilde{n}=0$ (infinite $\widetilde{v}$) corresponding to $\omega_{\mathrm{o}}\varphi_{\mathrm{o}}^{(1)}=\cot\varphi_{\mathrm{o}}$. This occupies a major portion of the upper right quadrant of Fig.~\ref{fig:GroupVelocity}(a) for normal AD.

\textit{Negative-$\widetilde{v}$ regime, $\widetilde{v}<0$ $(\widetilde{n}<0)$.} Negative values for $\widetilde{v}$ are achieved when $\omega_{\mathrm{o}}\varphi_{\mathrm{o}}^{(1)}\tan\varphi_{\mathrm{o}}>1$; for $\varphi_{\mathrm{o}}>0$, this requires that $\omega_{\mathrm{o}}\varphi_{\mathrm{o}}^{(1)}>0$ (anomalous AD).

\begin{figure}[t!]
\centering
\includegraphics[width=8.6cm]{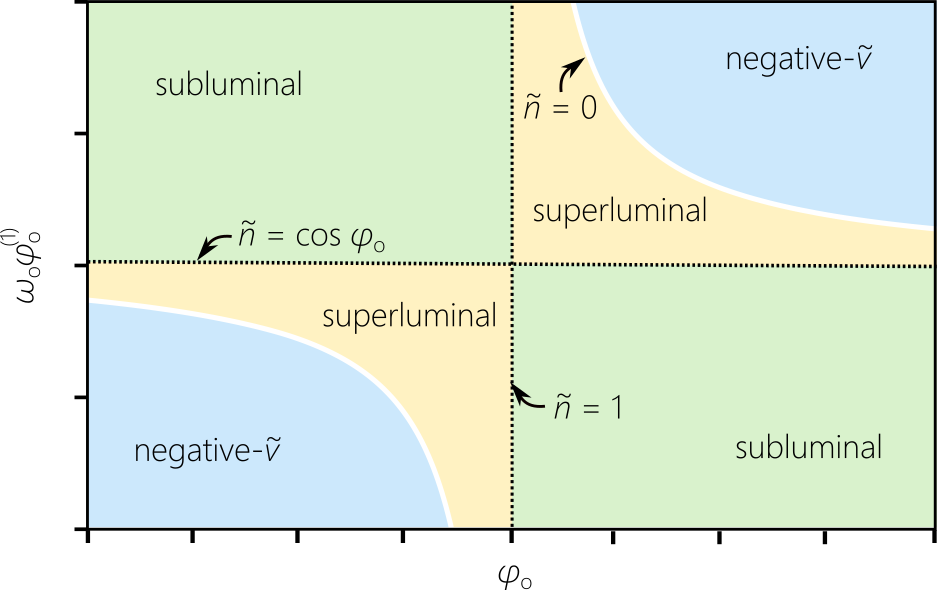}
\caption{Summary of the various regimes for tuning the AD-induced group velocity $\widetilde{v}$.}
\label{fig:GroupVelocitySummary}
\end{figure}

\textit{Summary of the AD-induced tunability of the group-velocity.} We identify the distinct regimes for tuning $\widetilde{v}$ in the simplified map illustrated in Fig.~\ref{fig:GroupVelocitySummary}. For a fixed positive angle $\varphi_{\mathrm{o}}>0$, $\widetilde{v}$ is always subluminal ($\widetilde{n}>1$) for $\omega_{\mathrm{o}}\varphi_{\mathrm{o}}^{(1)}\rightarrow-\infty$. Increasing $\omega_{\mathrm{o}}\varphi_{\mathrm{o}}^{(1)}$ in turn increases $\widetilde{v}$ (reduces $\widetilde{n}$) until we reach $\omega_{\mathrm{o}}\varphi_{\mathrm{o}}^{(1)}=-\tan\tfrac{\varphi_{\mathrm{o}}}{2}$, whereupon $\widetilde{v}=c$ is luminal ($\widetilde{n}=1$); see Fig.~\ref{fig:GroupVelocity}(b). When $\omega_{\mathrm{o}}\varphi_{\mathrm{o}}^{(1)}>-\tan\tfrac{\varphi_{\mathrm{o}}}{2}$, $\widetilde{v}$ becomes superluminal ($\widetilde{n}<1$) until $\omega_{\mathrm{o}}\varphi_{\mathrm{o}}^{(1)}=\cot\varphi_{\mathrm{o}}$, whereupon $\widetilde{n}=0$, after which we have negative-$\widetilde{v}$ ($\widetilde{n}<0$), which continues for all $\omega_{\mathrm{o}}\varphi_{\mathrm{o}}^{(1)}>\cot\varphi_{\mathrm{o}}$.

Alternatively, consider a fixed AD $\omega_{\mathrm{o}}\varphi_{\mathrm{o}}^{(1)}$ while scanning the angle $\varphi_{\mathrm{o}}\geq0$. First, take $\omega_{\mathrm{o}}\varphi_{\mathrm{o}}^{(1)}<0$ (normal AD), corresponding to the bottom right quadrant of Fig.~\ref{fig:GroupVelocity}(a). Starting at $\varphi_{\mathrm{o}}=0$, $\widetilde{v}$ is luminal ($\widetilde{n}=1$), after which $\widetilde{v}$ is subluminal ($\widetilde{n}>1$) as $\varphi_{\mathrm{o}}$ increases until we reach $\tan\tfrac{\varphi_{\mathrm{o}}}{2}=-\omega_{\mathrm{o}}\varphi_{\mathrm{o}}^{(1)}$, whereupon $\widetilde{v}$ is again luminal ($\widetilde{n}=1$). After this value, $\tan\tfrac{\varphi_{\mathrm{o}}}{2}>-\omega_{\mathrm{o}}\varphi_{\mathrm{o}}^{(1)}$, $\widetilde{v}$ is superluminal ($\widetilde{n}<1$) thereafter. When $|\omega_{\mathrm{o}}\varphi_{\mathrm{o}}^{(1)}|\geq1$ for normal AD, the wave packet remains subluminal for all values of $\varphi_{\mathrm{o}}$. Consider next $\omega_{\mathrm{o}}\varphi_{\mathrm{o}}^{(1)}>0$ (anomalous AD), corresponding to the top right quadrant of Fig.~\ref{fig:GroupVelocity}(a). Starting at $\varphi_{\mathrm{o}}=0$, $\widetilde{v}$ is again luminal ($\widetilde{n}=1$), after which $\widetilde{v}$ is superluminal ($\widetilde{n}<1$) with increasing $\varphi_{\mathrm{o}}$ until $\cot\varphi_{\mathrm{o}}=\omega_{\mathrm{o}}\varphi_{\mathrm{o}}^{(1)}$, whereupon $\widetilde{n}=0$. After this value of $\varphi_{\mathrm{o}}$, $\widetilde{v}$ is negative ($\widetilde{n}<0$).

\subsection{AD-induced GVD}

The quadratic $\Omega^{2}$~term in Eq.~\ref{eq:AxialWaveNumber} makes clear that AD introduces GVD even in free space. The AD-induced GVD coefficient~$k_{2}$ is given by:
\begin{equation}\label{eq:AD_GVD}
c\omega_{\mathrm{o}}k_{2}=-\left(2\omega_{\mathrm{o}}\varphi_{\mathrm{o}}^{(1)}+\omega_{\mathrm{o}}^{2}\varphi_{\mathrm{o}}^{(2)}\right)\sin\varphi_{\mathrm{o}}-\left(\omega_{\mathrm{o}}\varphi_{\mathrm{o}}^{(1)}\right)^{2}\cos\varphi_{\mathrm{o}}.
\end{equation}
Conventionally, $k_{2}>0$ is referred to as `normal GVD' and $k_{2}<0$ as `anomalous GVD' \cite{SalehBook07}. As a point of reference, GVD in silica glass at 1.5~$\mu$m is $c\omega_{\mathrm{o}}k_{2}\sim-0.01$. For on-axis fields ($\varphi_{\mathrm{o}}=0$), Eq.~\ref{eq:AD_GVD} simplifies to $c\omega_{\mathrm{o}}k_{2}=-\left(\omega_{\mathrm{o}}\varphi_{\mathrm{o}}^{(1)}\right)^{2}$, so that AD always yields anomalous GVD in free space, which implies that AD can be exploited for GVD cancellation in a material in the normal GVD regime. This result was formulated long ago by Martinez, Gordon, and Fork \cite{Martinez84JOSAA}. Nevertheless, independent control over $\varphi_{\mathrm{o}}$, $\varphi_{\mathrm{o}}^{(1)}$, and $\varphi_{\mathrm{o}}^{(2)}$ can flip the sign of GVD as suggested theoretically in Ref.~\cite{Porras03PRE2}. However, because such control has \textit{not} been forthcoming in conventional optics, neither AD-induced anomalous GVD in free space nor GVD cancellation in a medium in its anomalous GVD regime have been reported using conventional approaches; see, however, Ref.~\cite{Hall21OLNormalGVD} where a universal AD synthesizer was utilized to achieve this feat for the first time.

AD-induced GVD is a disadvantage when the goal is to tune $\widetilde{v}$. Any point in the graph in Fig.~\ref{fig:GroupVelocity}(a) corresponds to a field accompanied by GVD (Eq.~\ref{eq:AD_GVD}), leading to pulse dispersion and thus setting an upper limit on the propagation distance over which the group velocity $\widetilde{v}$ can be reliably observed \cite{Hall25APLP}. Nevertheless, AD-induced GVD is absent ($k_{2}=0$) when \cite{Porras03PRE2,Zapata06OL}:
\begin{equation}
\omega_{\mathrm{o}}^{2}\varphi_{\mathrm{o}}^{(2)}=-\omega_{\mathrm{o}}\varphi_{\mathrm{o}}^{(1)}\left\{2+\frac{\omega_{\mathrm{o}}\varphi_{\mathrm{o}}^{(1)}}{\tan\varphi_{\mathrm{o}}}\right\}.
\end{equation}
Eliminating AD-induced GVD while tuning $\widetilde{v}$ therefore requires independent control over $\varphi_{\mathrm{o}}$, $\varphi_{\mathrm{o}}^{(1)}$, and $\varphi_{\mathrm{o}}^{(2)}$. As we show in Part~II~\cite{Hall26JOSAA2}, an added advantage of non-differentiable AD is to eliminate GVD without impacting the tunability of $\widetilde{v}$.

\subsection{Practical limitations}

It is straightforward to introduce AD into a collimated polychromatic field using either a diffraction grating or prism (more recently, metasurfaces have been utilized). These devices allow for tuning the first-order AD coefficient $\omega_{\mathrm{o}}\varphi_{\mathrm{o}}^{(1)}$. However, such traditional optical devices do \textit{not} provide independent control over $\varphi_{\mathrm{o}}$ and $\varphi_{\mathrm{o}}^{(1)}$, and thus do not control the group velocity over the full plane depicted in Fig.~\ref{fig:GroupVelocity}(a).

\begin{figure*}[t!]
\centering
\includegraphics[width=17.6cm]{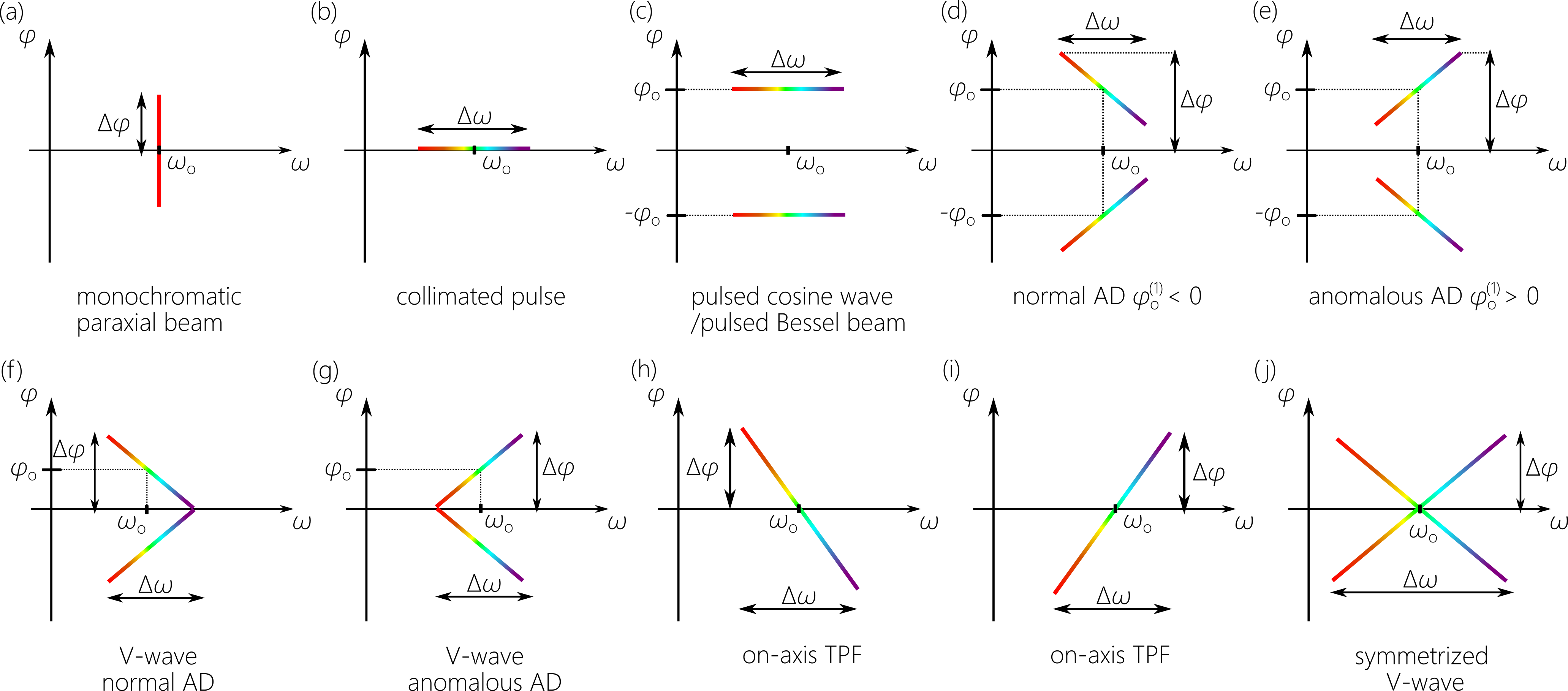}
\caption{Selecting $\varphi_{\mathrm{o}}$ for AD-endowed wave packets. (a) A monochromatic, on-axis paraxial beam at $\omega=\omega_{\mathrm{o}}$ with $\varphi_{\mathrm{o}}=0$. (b) A collimated, on-axis pulse with $\varphi(\omega)=\varphi_{\mathrm{o}}=0$. (c) Off-axis AD-free field, $\varphi(\omega)=\varphi_{\mathrm{o}}\neq0$ but $\varphi_{\mathrm{o}}^{(1)}=0$, corresponding to a pulsed cosine wave in the Cartesian AD plane, or a pulsed Bessel beam in the polar AD plane. (d, e) Off axis, AD-endowed  fields with (d) normal AD $\varphi_{\mathrm{o}}^{(1)}<0$ and (e) anomalous AD $\varphi_{\mathrm{o}}^{(1)}>0$. (f, g) Same as (d, e), except that $\varphi_{\mathrm{o}}=\tfrac{1}{2}\Delta\varphi$, so that the AD profiles reach $\varphi=0$ at one edge of the spectrum. Both of these cases correspond to so-called `V-waves' \cite{Hall21PRAVwave}. (h,i) Titled pulse fronts (TPFs). (j) A symmetrized V-wave.}
\label{fig:CentralFreqAngle}
\end{figure*}

Consider a diffraction grating normal to the $z$-axis, of period $\Lambda$, with light of central wavelength $\lambda_{\mathrm{o}}$ incident at an angle $\alpha$ with respect to the $z$-axis. The $m^{\mathrm{th}}$ order is diffracted at an angle $\varphi_{\mathrm{o}}$ with the $z$-axis, where $\sin\varphi_{\mathrm{o}}=m\tfrac{\lambda_{\mathrm{o}}}{\Lambda}+\sin\alpha$ \cite{SalehBook07}. The two independent parameters $\alpha$ and $m\tfrac{\lambda_{\mathrm{o}}}{\Lambda}$ determine $\varphi_{\mathrm{o}}$, from which we obtain the AD coefficient: $\omega_{\mathrm{o}}\varphi_{\mathrm{o}}^{(1)}=-m\tfrac{\lambda_{\mathrm{o}}}{\Lambda}\;\tfrac{1}{\cos\varphi_{\mathrm{o}}}$. Consequently, $\varphi_{\mathrm{o}}$ and $\varphi_{\mathrm{o}}^{(1)}$ are \textit{not} independently tunable. Moreover, $\varphi_{\mathrm{o}}^{(1)}$ is largest for large $\varphi_{\mathrm{o}}$ and $\alpha$. For example, normal incidence $\alpha=0$ on a grating with 1000~lines/mm produces a first diffraction order ($m=1$) with $\omega_{\mathrm{o}}\varphi_{\mathrm{o}}^{(1)}\approx -1.33$ at $\lambda_{\mathrm{o}}=0.8$~$\mu$m and $\varphi_{\mathrm{o}}\approx53^{\circ}$. Increasing the AD to $\omega_{\mathrm{o}}\varphi_{\mathrm{o}}^{(1)}=3.5$ requires increasing the incident angle to $10^{\circ}$, whereupon the diffraction angle increases to $\varphi_{\mathrm{o}}\approx76^{\circ}$.

Excluding glancing-angle incidence and diffraction, we have the approximate bound $|\omega_{\mathrm{o}}\varphi_{\mathrm{o}}^{(1)}|\lessapprox5$ \cite{Hall25APLP}, which restricts us to only a small portion of the plot in Fig.~\ref{fig:GroupVelocity}(a); see the expanded view in Fig.~\ref{fig:GroupVelocity}(c). Recent progress in the area of metasurfaces indicates the possibility of exercising independent control over multiple AD orders \cite{Arbabi2017Optica,Kamali2018Review,Wang2018Science,Qiu18PRAppl,McClung2020Light,Zhang20LSA,Li2021SciAdv}, but more work is needed along these lines. Although no one single optical \textit{device} to date provides full control over the AD profile, a novel \textit{system} that we call a `universal AD synthesizer' can produce arbitrary AD profiles. This system comprises a diffraction grating and spectral phase modulation \cite{Kondakci17NP,Kondakci18OE}, which together constitute a universal AD synthesizer along one transverse dimension \cite{Hall24JOSAA,Romer25JOpt}. Furthermore, this system can be rendered compact \cite{Yessenov23OL} by replacing the surface grating with a novel spectral analysis device: a rotated chirped volumetric Bragg grating (r-CBG) \cite{Mhibik23OL,Mhibik23OL2}. Tuning the AD profile arbitrarily in a circularly symmetric field requires a more sophisticated approach \cite{Yessenov22NC,Yessenov22OL,Yessenov25JOSAA} (some progress has been made in this regard using circular diffraction gratings \cite{Piccardo23NP}). It remains an open question whether a single optical device can provide access to independent control over $\varphi_{\mathrm{o}}$ and $\varphi_{\mathrm{o}}^{(1)}$ \cite{Hall25APLP} and thus tune the group velocity over the entire plane in Fig.~\ref{fig:GroupVelocity}(a).

\section{Defining the central propagation angle}

In most cases, the distinction between on-axis ($\varphi_{\mathrm{o}}=0$) and off-axis ($\varphi_{\mathrm{o}}\neq0$) fields is clear. Nevertheless, there are a few cases where ambiguity may arise, which we attempt to clarify here.

We provide some guidelines regarding the determination of $\varphi_{\mathrm{o}}$ for AD-endowed wave packets. We first define the temporal bandwidth $\Delta\omega$ and the angular bandwidth $\Delta\varphi$, the latter of which extends from the lowest to highest positive-valued angles in the Cartesian AD plane associated with the spectrum $\Delta\omega$ (and similarly for the polar AD plane). We assume throughout that $\Delta\omega\ll\omega_{\mathrm{o}}$, corresponding to the slowly varying temporal envelope approximation. Therefore, in general, when $\Delta\varphi\ll\varphi_{\mathrm{o}}$ we are in the off-axis regime.

For symmetrized AD-endowed fields along the $x$-axis, the field takes the form $E(x,z;t)=e^{ik_{\mathrm{o}}(z\cos\varphi_{\mathrm{o}}-ct)}\psi(x,z;t)$, where:
\begin{equation}
\psi(x,z;t)=\!\int\!d\Omega\widetilde{\psi}(\Omega)e^{-i\Omega(t-z/\widetilde{v})}\cos\{(k_{\mathrm{o}}\sin\varphi_{\mathrm{o}}+\widetilde{n}_{x}\tfrac{\Omega}{c})x\},
\end{equation}
where $\widetilde{v}=c/\widetilde{n}$, $\widetilde{v}$ is given by Eq.~\ref{eq:GroupVelocityGeneral}, and we have introduced a transverse group index:
\begin{equation}
\widetilde{n}_{x}=\sin\varphi_{\mathrm{o}}+\omega_{\mathrm{o}}\varphi_{\mathrm{o}}^{(1)}\cos\varphi_{\mathrm{o}}.
\end{equation}
In the polar AD plane, the envelope is correspondingly:
\begin{equation}\label{eq:SymmetrizedPolarWPs}
\psi(r,z;t)=\!\int\!d\Omega\widetilde{\psi}(\Omega)e^{-i\Omega(t-z/\widetilde{v})}J_{0}(\{k_{\mathrm{o}}\sin\varphi_{\mathrm{o}}+\widetilde{n}_{x}\tfrac{\Omega}{c}\}r).
\end{equation}
Note that $\widetilde{n}^{2}+\widetilde{n}_{x}^{2}=1+(\omega_{\mathrm{o}}\varphi_{\mathrm{o}}^{(1)})^{2}$, which becomes $\widetilde{n}^{2}+\widetilde{n}_{x}^{2}=1$ in absence of AD. That is, in absence of AD, the group index is always superluminal $\widetilde{n}=\cos\varphi_{\mathrm{o}}<1$, which is associated with a superluminal transverse group index $\widetilde{n}_{x}=\sin\varphi_{\mathrm{o}}<1$. The presence of AD allows for other regimes to arise [Fig.~\ref{fig:GroupVelocity}(a)].

We show in Fig.~\ref{fig:CentralFreqAngle} the most salient examples of AD profiles. In Fig.~\ref{fig:CentralFreqAngle}(a) we depict a monochromatic paraxial beam at a fixed frequency $\omega=\omega_{\mathrm{o}}$, so that $E(x,z;t)=e^{ik_{\mathrm{o}}(z-ct)}\psi(x,z)$, where $\psi(x,z)$ is the usual diffracting spatial profile, the phase velocity is $v_{\mathrm{ph}}=c$, and the group velocity is not defined for a monochromatic beam. By contrast, Fig.~\ref{fig:CentralFreqAngle}(b) depicts a collimated pulse with $k_{z}(\omega)=\tfrac{\omega}{c}$, $k_{x}=0$, $\varphi(\omega)=0$, $\varphi_{\mathrm{o}}^{(1)}=0$, $\widetilde{n}=1$, and $\widetilde{n}_{x}=0$, so that $E(x,z;t)=e^{ik_{\mathrm{o}}(z-ct)}\psi(z;t)$. Here we have $v_{\mathrm{ph}}=\widetilde{v}=c$, $\psi(z;t)=\psi(0,t-z/c)$, and $\psi(0;t)=\int\!d\Omega\;\widetilde{\psi}(\Omega)e^{-i\Omega t}$. As expected, for a collimated on-axis pulse in free space the group velocity is $\widetilde{v}=c$. Displacing this AD-free pulse to off-axis propagation with $\varphi_{\mathrm{o}}\neq0$ and $\varphi_{\mathrm{o}}^{(1)}=0$ [Fig.~\ref{fig:CentralFreqAngle}(c) and Fig.~\ref{fig:PhaseAndGroup}(b)], yields a symmetrized wave packet with $v_{\mathrm{ph}}=\widetilde{v}=c/\cos\varphi_{\mathrm{o}}$, $\widetilde{n}=\cos\varphi_{\mathrm{o}}$, $\widetilde{n}_{x}=\sin\varphi_{\mathrm{o}}$, and:
\begin{equation}
\psi(x,z;t)=\!\!\int\!\!d\Omega\;\widetilde{\psi}(\Omega)e^{-i\Omega(t-z/\widetilde{v})}\cos\left\{\tfrac{\omega_{\mathrm{o}}+\Omega}{c}\sin\varphi_{\mathrm{o}}x\right\}.
\end{equation}

In Fig.~\ref{fig:CentralFreqAngle}(d,e) we depict off-axis fields endowed with normal and anomalous AD. In both cases, $\Delta\varphi\ll\varphi_{\mathrm{o}}$, so that they are unambiguously off-axis fields. Along the $z$-axis we have $v_{\mathrm{ph}}=c/\cos\varphi_{\mathrm{o}}$, while $\widetilde{v}$ is given by Eq.~\ref{eq:GroupVelocityGeneral}. The spectrum is centered at $\omega_{\mathrm{o}}$, which is associated with the central propagation angle $\varphi_{\mathrm{o}}$. Strictly speaking, $\varphi_{\mathrm{o}}$ will not be the central angle unless $\varphi(\omega)$ is linear in $\omega$. If only small departures from linearity occur, then this deviation can be ignored. Indeed, the useful spectrum in many cases is usually taken to be restricted to the linear AD regime.

The scenarios depicted in Fig.~\ref{fig:CentralFreqAngle}(f-i) are borderline cases and require care. We first depict in Fig.~\ref{fig:CentralFreqAngle}(f,g) the scenarios where the central propagation angle is reduced so that $\varphi_{\mathrm{o}}=\tfrac{1}{2}\Delta\varphi$ and the angular spectrum reaches $\varphi(\omega_{\mathrm{o}}+\tfrac{\Delta\omega}{2})=0$ at the edge of the spectrum in Fig.~\ref{fig:CentralFreqAngle}(f) and $\varphi(\omega_{\mathrm{o}}-\tfrac{\Delta\omega}{2})=0$ in Fig.~\ref{fig:CentralFreqAngle}(g). This borderline case still strictly represents an off-axis field with $\widetilde{n}=\cos\varphi_{\mathrm{o}}\pm2\tfrac{\varphi_{\mathrm{o}}}{\Delta\omega/\omega_{\mathrm{o}}}\sin\varphi_{\mathrm{o}}$; here, the positive sign corresponds to Fig.~\ref{fig:CentralFreqAngle}(f) and the negative to Fig.~\ref{fig:CentralFreqAngle}(g). Because of the shape of their angular spectra (when $\omega$ is taken to be the vertical axis), we have called such fields V-waves \cite{Hall21PRAVwave}. Except for the pulse-front tilt and the anomalous AD-induced GVD that characterize this wave packet \cite{Hall21PRAVwave}, the group velocity approaches $c$ for small $\varphi_{\mathrm{o}}$.

The two examples in Fig.~\ref{fig:CentralFreqAngle}(h,i) represent on-axis, AD-endowed fields with $v_{\mathrm{ph}}=\widetilde{v}=c$ along the $z$-axis. Each is simply a TPF \cite{Fulop10Review} that does \textit{not} correspond to a symmetrized field and does not have an on-axis intensity peak. Both of these TPFs experience AD-induced anomalous GVD in free space \cite{Porras03PRE2,Martinez84JOSAA}. These two on-axis TPFs cannot be directly assigned to normal or anomalous AD. Indeed, each case is a combination of both. For example, in Fig.~\ref{fig:CentralFreqAngle}(h), the portion $\varphi>0$ of the AD spectrum is associated with normal AD, whereas the remaining portion $\varphi<0$ is associated with anomalous AD and vice versa for the TPF in Fig.~\ref{fig:CentralFreqAngle}(i).

We can form a symmetrized wave packet by combining the AD profiles for the V-waves in Fig.~\ref{fig:CentralFreqAngle}(h,i), and the corresponding AD profile is plotted in Fig.~\ref{fig:CentralFreqAngle}(j), which is therefore a symmetrized V-wave with $\varphi_{\mathrm{o}}=0$ and $v_{\mathrm{ph}}=\widetilde{v}=c$. This wave packet also undergoes AD-induced anomalous GVD in free space. The symmetrized V-wave in Fig.~\ref{fig:CentralFreqAngle}(j) can also be viewed as concatenating the two V-waves in Fig.~\ref{fig:CentralFreqAngle}(f,g).


\section{Special cases for tuning the group velocity via angular dispersion}

\subsection{Fixed propagation angle: X-waves}

\begin{figure}[t!]
\centering
\includegraphics[width=8.8cm]{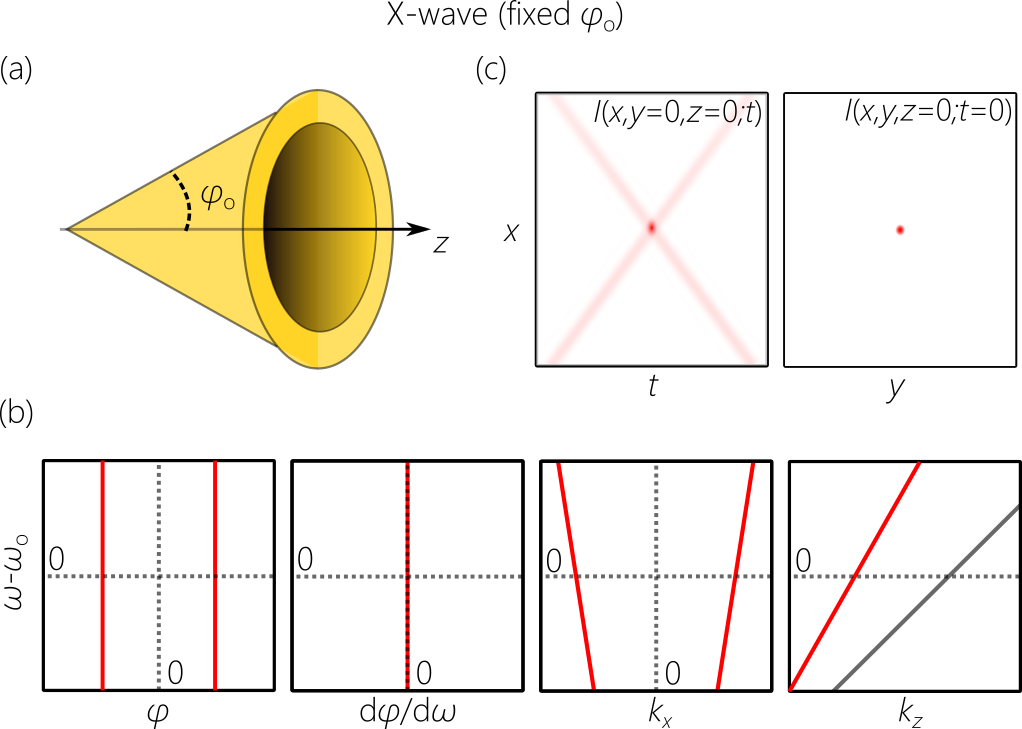}
\caption{X-waves having a fixed propagation angle $\varphi(\omega)=\varphi_{\mathrm{o}}$. (a) A pictorial depiction of the conical AD for an X-wave. (b) Sketches of $\varphi(\omega)$, $\tfrac{d\varphi}{d\omega}$, $k_{x}(\omega)$ and $k_{z}(\omega)$. For convenience, we plot using $\omega$ as the vertical axis. The black line in the last panel is the free-space light-line $k_{z}=\tfrac{\omega}{c}$. (c) A section through the spatiotemporal intensity $I(x,y=0,z=0;t)$ of an X-wave at a fixed axial plane $z=0$, in a meridional plane containing the $z$-axis ($y=0$); and the transverse spatial profile at a fixed axial plane $z=0$ and fixed time $t=0$, $I(x,y,z=0;t=0)$.}
\label{fig:BesselX}
\end{figure}

Consider an azimuthally symmetric, off-axis, AD-free wave packet in which all the frequencies travel at \textit{the same angle}, $\varphi(\omega)=\varphi_{\mathrm{o}}\neq0$, and all AD coefficients vanish [Fig.~\ref{fig:BesselX}(a)], $\varphi_{\mathrm{o}}^{(n)}\!=\!0$ ($n\geq1$), in which case $\widetilde{n}=\cos\varphi_{\mathrm{o}}$ and $\widetilde{n}_{x}=\sin\varphi_{\mathrm{o}}$  [Fig.~\ref{fig:BesselX}(b)], the phase and group velocities are equal:
\begin{equation}
\widetilde{v}=v_{\mathrm{ph}}=\frac{c}{\cos{\varphi_{\mathrm{o}}}}>c,
\end{equation}
which is always \textit{superluminal} ($\widetilde{v}\!\rightarrow\!c$ when $\varphi_{\mathrm{o}}\!\rightarrow\!0$). The field is:
\begin{equation}
E(r,z;t)=\int\!d\omega\;\widetilde{\psi}(\omega)\;J_{0}(\tfrac{\omega}{c}r\sin\varphi_{\mathrm{o}})e^{i\omega(t-z/\widetilde{v})},
\end{equation}
with no phase external to the integral representing a carrier. This class of wave packets is known as X-waves \cite{Lu92IEEEa,Lu92IEEEb,Saari97PRL,Yessenov19PRA}; see Fig.~\ref{fig:BesselX}(c). Because $E(x,z;t)=E(x,0;t-z/\widetilde{v})$, X-waves are propagation invariant \cite{FigueroaBook14}; that is, the spatiotemporal profile travels rigidly in free space (without diffraction or AD-induced dispersion) at a fixed group velocity $\widetilde{v}$.

The deviation in $\widetilde{v}$ from $c$ for this AD-free wave packet stems from a purely geometric effect ($\varphi_{\mathrm{o}}\neq0$). Only minute deviations in $\widetilde{v}$ can thus be realized for an X-wave by changing $\varphi_{\mathrm{o}}$ in the paraxial (or small-angle) regime; e.g., at $\varphi_{\mathrm{o}}=1^{\circ}$ we have $\widetilde{v}=1.00015c$, and when $\varphi_{\mathrm{o}}=5^{\circ}$ we have $\widetilde{v}=1.0038c$. All experimentally reported values of the group velocity of X-waves have been close to $c$: $\tilde{v} \approx 1.00022c$ \cite{Bonaretti09OE}, $\tilde{v} \approx 1.00015c$ \cite{Kuntz09PRA}, and $\tilde{v} \approx 1.00012c$ \cite{Bowlan09OL}. A substantial deviation in $\widetilde{v}$ requires operating deep in the non-paraxial regime; e.g., reaching $\widetilde{v}=1.1c$ requires $\varphi_{\mathrm{o}}\approx24.6^{\circ}$, and $\widetilde{v}=2c$ requires $\varphi_{\mathrm{o}}=60^{\circ}$.

\subsection{Fixed transverse wave number: Pulsed Bessel beams}

\begin{figure}[t!]
\centering
\includegraphics[width=8.8cm]{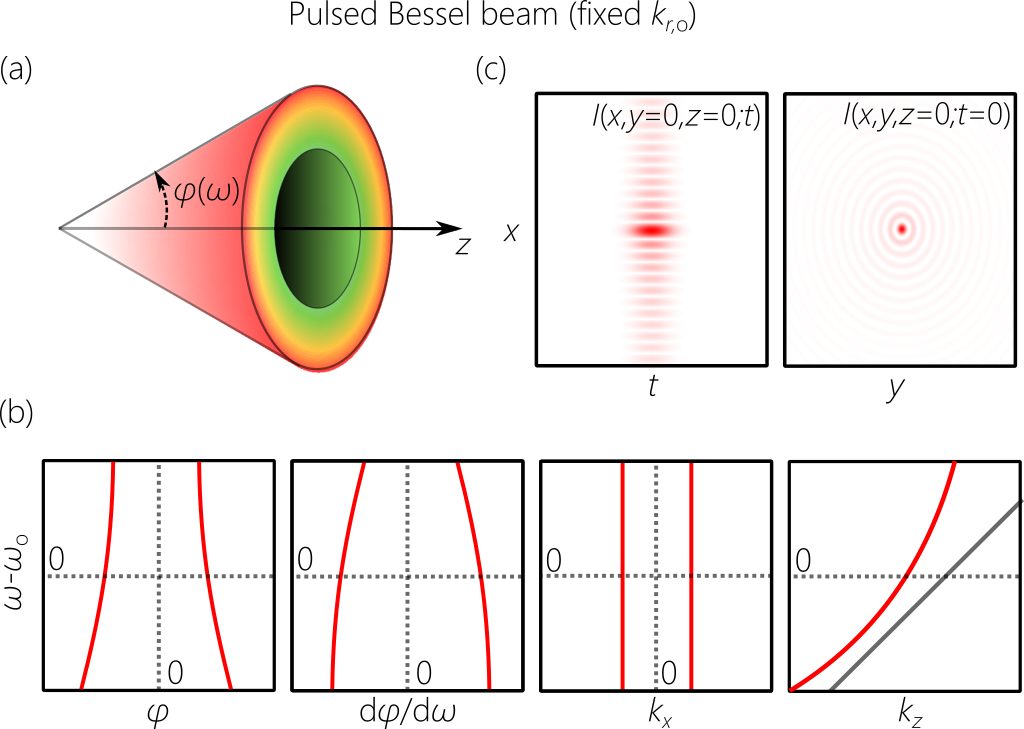}
\caption{Same as Fig.~\ref{fig:BesselX} for a pulsed Bessel beam having a fixed radial wave number $k_{r}=k_{r,{\mathrm{o}}}$.}
\label{fig:PulsedBesselBeam}
\end{figure}

Consider alternatively an azimuthally symmetric wave packet in which the transverse wave number $k_{r}(\omega)=\tfrac{\omega}{c}\sin\varphi(\omega)=k_{r,\mathrm{o}}$ is held fixed independently of $\omega$ [Fig.~\ref{fig:PulsedBesselBeam}(a,b)], whereupon $\omega_{\mathrm{o}}\varphi_{\mathrm{o}}^{(1)}=-\tan\varphi_{\mathrm{o}}$, $\widetilde{n}_{x}=0$, and $\omega_{\mathrm{o}}^{2}\varphi_{\mathrm{o}}^{(2)}=\tan\varphi_{\mathrm{o}}(2-\tan^{2}\varphi_{\mathrm{o}})$. Along the $z$-axis, $v_{\mathrm{ph}}=c/\cos\varphi_{\mathrm{o}}$ and $\widetilde{v}$ is:
\begin{equation}
\widetilde{v}=c\cos\varphi_{\mathrm{o}}<c,
\end{equation}
which is always \textit{subluminal}, and can be tuned by varying $\varphi_{\mathrm{o}}$. The field envelope separates into a product of a transverse spatial profile (a diffraction-free Bessel beam) and an axially evolving temporal profile [Fig.~\ref{fig:PulsedBesselBeam}(c)]:
\begin{equation}
\psi(r,z;t)=J_{0}(k_{\mathrm{o}}r\sin\varphi_{\mathrm{o}})\psi(z;t),
\end{equation}
where $\psi(z;t)=\int\!d\Omega\;\widetilde{\psi}(\Omega)e^{-i\Omega(t-z/\widetilde{v})}e^{ik_{2}\Omega^{2}z/2}$, and $c\omega_{\mathrm{o}}k_{2}=\tan\varphi_{\mathrm{o}}\sin\varphi_{\mathrm{o}}\left(\tan^{2}\varphi_{\mathrm{o}}-1\right)$ [Fig.~\ref{fig:PulsedBesselBeam}(b)]. This can be realized by imposing a fixed spatial profile $J_{0}(k_{r,\mathrm{o}}r)$ on all wavelengths. Such a wave packet is usually called a pulsed Bessel beam \cite{Campbell90JASA,Liu98JMO,Hu02JOSAA,Lu03JOSAA,Turunen10PO}. In this field configuration, $\widetilde{v}v_{\mathrm{ph}}\!=\!c^{2}$ (similarly to an optical waveguide with perfect mirror boundaries \cite{SalehBook07}).

Pulsed Bessel beams suffer from the same difficulty as X-waves: a large deviation in $\widetilde{v}$ requires operating in the non-paraxial regime. Whereas $\widetilde{v}=0.99985c$ requires $\varphi_{\mathrm{o}}=1^{\circ}$ and $\widetilde{v}=0.9962c$ requires $\varphi_{\mathrm{o}}=5^{\circ}$, to reach $\widetilde{v}=0.9c$ requires $\varphi_{\mathrm{o}}\approx25.84^{\circ}$ and $\widetilde{v}=0.5c$ requires $\varphi_{\mathrm{o}}=60^{\circ}$, both of which are deeply non-paraxial. For small angles we have $\omega_{\mathrm{o}}\varphi_{\mathrm{o}}^{(1)}\approx-\varphi_{\mathrm{o}}$ and $c\omega_{\mathrm{o}}k_{2}\approx-\varphi_{\mathrm{o}}^{2}$, so that GVD can usually be neglected. Increasing $\varphi_{\mathrm{o}}$ to change $\widetilde{v}$ significantly in the off-axis regime further suffers from increased GVD. For example, at $\varphi_{\mathrm{o}}=25.84^{\circ}$ ($\widetilde{v}=0.9c$) we have $c\omega_{\mathrm{o}}k_{2}\approx-0.16$, and at $\varphi_{\mathrm{o}}=60^{\circ}$ ($\widetilde{v}=0.5c$) we have $c\omega_{\mathrm{o}}k_{2}\approx3$, which are more than one and two orders-of-magnitude larger than silica glass at 1.5~$\mu$m, respectively. In Ref.~\cite{Giovannini15Science}, $\varphi_{\mathrm{o}}\approx0.26^{\circ}$, so that GVD is negligible over a 1-m propagation distance.

\subsection{Titled pulse front with negative group velocity}

A proposal for a negative-$\widetilde{v}$ AD-endowed tilted pulse front ($\omega_{\mathrm{o}}\varphi_{\mathrm{o}}^{(1)}\tan\varphi_{\mathrm{o}}>1$) was put forth by Zapata and Porras in Ref.~\cite{Zapata06OL}. To remain in the paraxial regime (small $\varphi_{\mathrm{o}}$), the required AD $\omega_{\mathrm{o}}\varphi_{\mathrm{o}}^{(1)}>\tfrac{1}{\tan\varphi_{\mathrm{o}}}$ is very large; e.g., setting $\varphi_{\mathrm{o}}=2^{\circ}$ necessitates $\omega_{\mathrm{o}}\varphi_{\mathrm{o}}^{(1)}>29^{\circ}$, which is an extremely large value. Furthermore, independent control over $\varphi_{\mathrm{o}}$ and $\varphi_{\mathrm{o}}^{(1)}$ is required. Such a wave packet is accompanied by GVD, unless $\varphi_{\mathrm{o}}^{(2)}$ can also be independently controlled. Setting $\varphi_{\mathrm{o}}=2^{\circ}$ and $\omega_{\mathrm{o}}\varphi_{\mathrm{o}}^{(1)}=40$, GVD-free propagation is achieved when $\omega_{\mathrm{o}}^{2}\varphi_{\mathrm{o}}^{(2)}\approx-140$, which is an extremely large value yet to be produced with conventional optics. The exorbitantly high AD values required, along with the necessary independent control over $\varphi_{\mathrm{o}}$, $\varphi_{\mathrm{o}}^{(1)}$, and $\varphi_{\mathrm{o}}^{(2)}$ prohibit the observation of negative-$\widetilde{v}$ in free space in the paraxial domain. These difficulties emphasize the tremendous prospects of non-differentiable AD that circumscribes all these challenges.

\section{Discussion}\label{sec:Discussion}

The notion of an optical wave packet having a superluminal group velocity is troubling because it appears to contradict the conventional understanding of relativistic causality. However, the group velocity is simply the speed of the wave-packet peak -- which does \textit{not} necessarily coincide with the energy velocity \cite{Smith70AJP} nor the information velocity \cite{Stenner2003Nature}. In general, relativistic causality does not constrain the group velocity \cite{Landauer89Nature,Landauer92SSC,Landauer93Nature,Diener96PLA,Chiao02OPN,Milonni02JPB,Milonni20Book}.

\begin{figure}[t!]
\centering
\includegraphics[width=8.8cm]{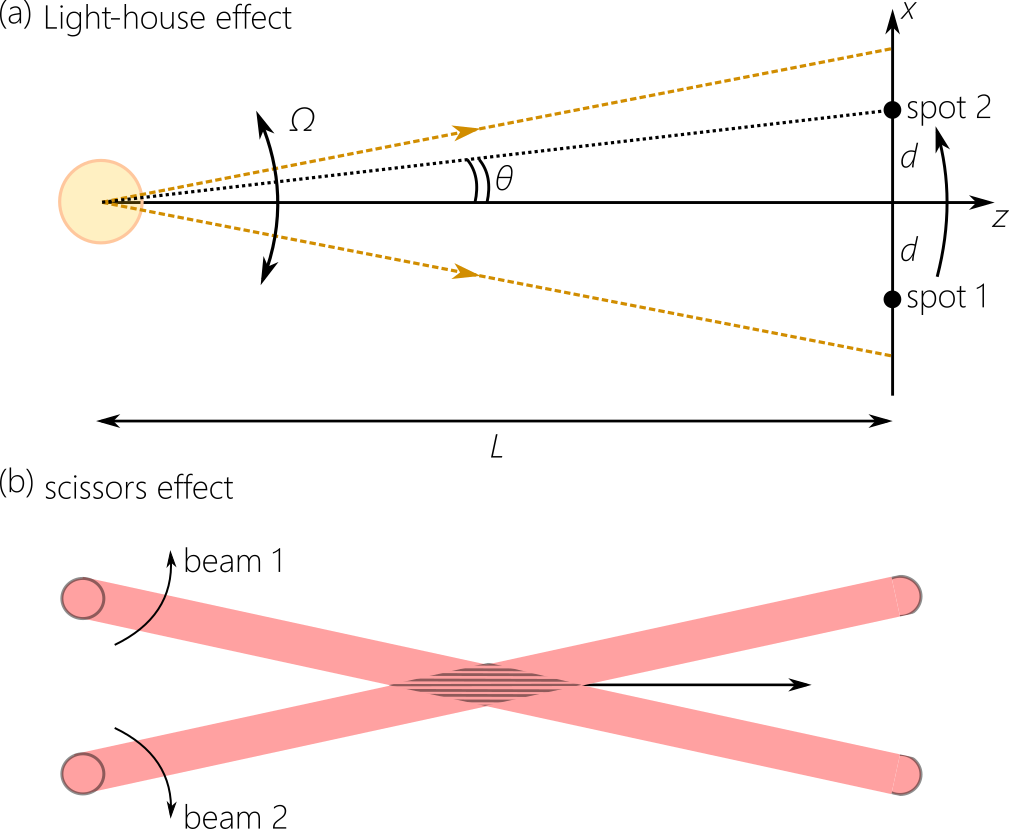}
\caption{(a) Illustration of the `light-house effect' in which apparent superluminal motion of an optical beam is produced through purely geometric means. (b) Illustration of the `scissors effect' produced by two intersecting beams of light. When the two beams are rotated away from each other, their intersection points moves forward and can become superluminal. This superluminal behavior is a purely geometrical effect.}
\label{fig:LightHouse}
\end{figure}

A standard thought experiment to produce `superluminal' light is that of an optical beam emitted by a rapidly rotating source and detected at a distant plane, sometimes known as the `light-house effect' \cite{Salmon16Book}; see Fig.~\ref{fig:LightHouse}(a). The beam travels at $c$ to the distant plane, but if the source rotates at a sufficiently rapid rate, the beam spot on the detection plane can move laterally along $x$ at a speed nominally higher than $c$. This of course poses no challenge to relativistic causality because any two spots on the target plane are not causally related and are instead tied to the source rather than to each other. No change at spot~1 is relayed over to spot~2, which requires instead a beam of light to be generated at spot~1 and sent to spot~2 (traveling at $c$ in free space). This effect is purely geometric and causes no confusion.

A numerical example helps illustrate this effect. Consider an observation plane at a distance $L$ from the source, with spot~1 and spot~2 at positions $d$ and $-d$ from the origin along the $x$-axis, respectively [Fig.~\ref{fig:LightHouse}(a)]. The source rotates at a rate $\Omega$~rad/s, and the beam moves along $x$ from spot~1 to spot~2 in a time interval of~$2T$. If $v=\tfrac{d}{T}>c$, then the beam appears to be superluminal along the $x$-axis at the observation plane. From the geometry of the problem we have $d=L\tan(\Omega T)$, so that the superluminal condition occurs whenever $\Omega>\tfrac{1}{T}\tan^{-1}(\tfrac{cT}{L})$. In the case where $\tfrac{cT}{L}\ll1$, we have the condition $\Omega>\tfrac{c}{L}$ for an apparently superluminal spot along $x$. Take $L=3\times10^{6}$~m, then $\Omega>100$~rad/s -- which is a readily achievable value -- is required for the beam to travel from spot~1 to spot~2 at a superluminal speed. Taking $\Omega=200$~rad/s leads to $v\approx2c$, and setting $T=0.1$~ms yields $d\approx60$~km ($\ll L$). 

A related thought experiment is that of a giant pair of scissors as it is being opened or closed. The moving intersection point of the two blades can move at a superluminal speed if the blades are opened at a sufficiently high (but subluminal) speed. An optical version of this involves two intersecting beams of light. Changing the propagation angles of the two beams in opposite directions in tandem can lead to their intersection moving superluminally [Fig.~\ref{fig:LightHouse}(b)]. Once again, this phenomenon can be attributed to a purely geometric effect. This class of effects are closely related to AD-free X-waves whose superluminal group velocity is indeed a consequence of purely geometric considerations. 

It is sometimes thought that all cases of superluminal light fall under this conceptual umbrella. However, this undercuts the role of interference in reshaping optical pulses in the time domain and optical beams in the spatial domain. In the case of `fast light', a superluminal group velocity is introduced into a collimated pulse by the chromatic dispersion accompanying an absorption resonance (close to the resonance frequency). The dispersion profile (the frequency-dependent refractive index) associated with such a resonance introduces a spectral phase that reshapes the pulse and moves its peak forward towards the pulse leading edge (but never exceeding it), accompanied by heavy losses. Conceptually, this phenomenon is not different from `slow light' except that the dispersion sign is reversed, leading to pulse reshaping in such a way that the pulse peak moves towards the trailing edge. Although both slow and fast light emerge in principle from a similar effect, fast light understandably appears more troubling than slow light.

In the case of AD-endowed wave packets, two effects determine the group velocity:
\begin{enumerate}
\item A geometric effect due to the central propagation angle $\varphi_{\mathrm{o}}$, which is reminiscent of the `light-house' or `scissor' effects.
\item Spatiotemporal interference associated with the linear AD term $\omega_{\mathrm{o}}\varphi_{\mathrm{o}}^{(1)}$.
\end{enumerate}
These two effects interact to yield the group velocity $\widetilde{v}$. The subluminal pulsed Bessel beam in Ref.~\cite{Giovannini15Science} is an example of such an effect. Even for a small propagation angle $\varphi_{\mathrm{o}}$, where the geometric effect becomes negligible, the group velocity can still be tuned by varying $\omega_{\mathrm{o}}\varphi_{\mathrm{o}}^{(1)}$. However, because it is difficult to tune $\varphi_{\mathrm{o}}$ and $\varphi_{\mathrm{o}}^{(1)}$ independently, such effects have not been examined systematically to date.

Consider the following numerical example for a pulsed Bessel beam. If $\varphi_{\mathrm{o}}=1^{\circ}$, then $\omega_{\mathrm{o}}\varphi_{\mathrm{o}}^{(1)}=-\tan\varphi_{\mathrm{o}}\approx-0.01746$ and $\widetilde{v}=c\cos\varphi_{\mathrm{o}}\approx0.99985c$. If one changes only the sign of the AD coefficient so that $\omega_{\mathrm{o}}\varphi_{\mathrm{o}}^{(1)}=\tan\varphi_{\mathrm{o}}\approx0.01746$, then we have $\widetilde{v}=c\tfrac{\cos\varphi_{\mathrm{o}}}{\cos2\varphi_{\mathrm{o}}}\approx1.0046c>c$. In other words, for the same propagation angle $\varphi_{\mathrm{o}}$, simply switching the sign of the AD coefficient can flip the group velocity regime from subluminal to superluminal -- just as switching from an absorption resonance to a gain resonance flips from the fast-light regime to the slow-light regime (in the vicinity of the resonance frequency). Although conventional systems for introducing AD do not usually provide this capability, the class of universal AD synthesizers in Refs.~\cite{Hall24JOSAA,Romer25JOpt,Yessenov25JOSAA} open the path for such investigations.

\section{Conclusion}

In conclusion, we have shown that the common intuition suggesting that spatially or spatiotemporally structuring an optical pulse always reduces its group velocity in free space below $c$ is \textit{not} entirely correct. Geometric \textit{and} interferometric effects combine to change the wave packet group velocity, potentially yielding arbitrary values in free space measured along a prescribed axis. We have focused on a field model in which the pulse is endowed with AD, whereby a fixed group velocity is maintained along the pulse propagation axis and over its cross section. The geometric effect arises from the propagation angle of the wave packet with respect to the optical axis, and the interferometric effect arises from the first-order AD coefficient (the rate of change of the propagation angle with respect to the optical frequency). We have found that significant deviation from $c$ in free space necessitates operating deep in the non-paraxial regime when relying on conventional (differentiable) AD. Moreover, a wave packet endowed with AD experiences AD-induced GVD in free space. We show in Part~II of this tutorial that the limits described here are circumvented by implementing the recently discovered form of AD referred to as non-differentiable AD. 

\medskip
\textbf{Funding} U.S. Office of Naval Research (ONR) N00014-19-1-2192 and N00014-20-1-2789. \\

\textbf{Disclosures} The authors declare no conflicts of interest. \\

\textbf{Data availability} The data that support the findings of this article are not publicly available. The data are available from the author upon reasonable request.

\bibliography{diffraction}

@BOOK{SalehBook07,
  AUTHOR =       {B. E. A. Saleh and M. C. Teich},
  TITLE =        {Fundamentals of Photonics},
  PUBLISHER =    {Wiley},
  YEAR =         {2007},
  address =      {},
}

@BOOK{FigueroaBook14,
  EDITOR =       {H. E. Hern\'andez-Figueroa and E. Recami and M. Zamboni-Rached},
  TITLE =        {Non-diffracting Waves},
  PUBLISHER =    {Wiley-VCH},
  YEAR =         {2014},
  address =      {},
}

@ARTICLE{Kondakci16OE,
  AUTHOR =       {H. E. Kondakci and A. F. Abouraddy},
  TITLE =        {Diffraction-free pulsed optical beams via space-time correlations},
  JOURNAL =      {Opt. Express},
  YEAR =         {2016},
  volume =       {24},
  pages =        {28659-28668},
}

@ARTICLE{Turunen10PO,
  AUTHOR =       {J. Turunen and A. T. Friberg},
  TITLE =        {Propagation-invariant optical fields},
  JOURNAL =      {Prog. Opt.},
  YEAR =         {2010},
  volume =       {54},
  pages =        {1-88},
}

@ARTICLE{Lu92IEEEa,
  AUTHOR =       {J.-Y. Lu and J. F. Greenleaf},
  TITLE =        {Nondiffracting {X} waves -- exact solutions to free-space scalar wave equation and their finite aperture realizations},
  JOURNAL =      {IEEE Trans. Ultrason. Ferroelec. Freq. Control},
  YEAR =         {1992},
  volume =       {39},
  pages =        {19-31},
}

@ARTICLE{Lu92IEEEb,
  AUTHOR =       {J.-Y. Lu and J. F. Greenleaf},
  TITLE =        {Experimental verification of nondiffracting {X} waves},
  JOURNAL =      {IEEE Trans. Ultrason. Ferroelec. Freq. Control},
  YEAR =         {1992},
  volume =       {39},
  pages =        {441-446},
}

@ARTICLE{Saari97PRL,
  AUTHOR =       {P. Saari and K. Reivelt},
  TITLE =        {Evidence of {X}-shaped propagation-invariant localized light waves},
  JOURNAL =      {Phys. Rev. Lett.},
  YEAR =         {1997},
  volume =       {79},
  pages =        {4135-4138},
}

@ARTICLE{Kondakci17NP,
  AUTHOR =       {H. E. Kondakci and A. F. Abouraddy},
  TITLE =        {Diffraction-free space-time light sheets},
  JOURNAL =      {Nat. Photon.},
  YEAR =         {2017},
  volume =       {11},
  pages =        {733-740},
}

@ARTICLE{Kondakci18OE,
  AUTHOR =       {H. E. Kondakci and M. Yessenov and M. Meem and D. Reyes and D. Thul and S. Rostami Fairchild and M. Richardson and R. Menon and A. F. Abouraddy},
  TITLE =        {Synthesizing broadband propagation-invariant space-time wave packets using transmissive phase plates},
  JOURNAL =      {Opt. Express},
  YEAR =         {2018},
  volume =       {26},
  pages =        {13628-13638},
}

@ARTICLE{Yessenov19PRA,
  AUTHOR =       {M. Yessenov and B. Bhaduri and H. E. Kondakci and A. F. Abouraddy},
  TITLE =        {Classification of propagation-invariant space-time light-sheets in free space: Theory and experiments},
  JOURNAL =      {Phys. Rev. A},
  YEAR =         {2019},
  volume =       {99},
  pages =        {023856},
}

@ARTICLE{Hu02JOSAA,
  AUTHOR =       {W. Hu and H. Guo},
  TITLE =        {Ultrashort pulsed {B}essel beams and spatially induced group-velocity dispersion},
  JOURNAL =      {J. Opt. Soc. Am. A},
  YEAR =         {2002},
  volume =       {19},
  pages =        {49-53},
}

@ARTICLE{Parker16OE,
  AUTHOR =       {K. J. Parker and M. A. Alonso},
  TITLE =        {The longitudinal iso-phase condition and needle pulses},
  JOURNAL =      {Opt. Express},
  YEAR =         {2016},
  volume =       {24},
  pages =        {28669-28677},
}

@ARTICLE{SaintMarie17Optica,
  AUTHOR =       {A. Sainte-Marie and O. Gobert and F. Qu{\'e}r{\'e}},
  TITLE =        {Controlling the velocity of ultrashort light pulses in vacuum through spatio-temporal couplings},
  JOURNAL =      {Optica},
  YEAR =         {2017},
  volume =       {4},
  pages =        {1298--1304},
}

@ARTICLE{Bhaduri19Optica,
  AUTHOR =       {B. Bhaduri and M. Yessenov and A. F. Abouraddy},
  TITLE =        {Space-time wave packets that travel in optical materials at the speed of light in vacuum},
  JOURNAL =      {Optica},
  YEAR =         {2019},
  volume =       {6},
  pages =        {139-146},
}

@ARTICLE{Lu03JOSAA,
  AUTHOR =       {B. L{\"u} and Z. Liu},
  TITLE =        {Propagation properties of ultrashort pulsed {B}essel beams in dispersive media},
  JOURNAL =      {J. Opt. Soc. Am. A},
  YEAR =         {2003},
  volume =       {20},
  pages =        {582-587},
}

@ARTICLE{Kondakci19NC,
  AUTHOR =       {H. E. Kondakci and A. F. Abouraddy},
  TITLE =        {Optical space-time wave packets of arbitrary group velocity in free space},
  JOURNAL =      {Nat. Commun.},
  YEAR =         {2019},
  volume =       {10},
  pages =        {929},
}

@ARTICLE{Kondakci19OL,
  AUTHOR =       {H. E. Kondakci and M. A. Alonso and A. F. Abouraddy},
  TITLE =        {Classical entanglement underpins the propagation invariance of space-time wave packets},
  JOURNAL =      {Opt. Lett.},
  YEAR =         {2019},
  volume =       {44},
  pages =        {2645-2648},
}

@Book{Salmon16Book,
  AUTHOR =       {W. C. Salmon},
  TITLE =        {Four Decades of Scientific Explanation},
  Publisher =    {Univ. Pittsburgh Press},
  YEAR =         {2006},
  }

@ARTICLE{Boyd09Science,
  AUTHOR =       {R. W. Boyd and D. J. Gauthier},
  TITLE =        {Controlling the velocity of light pulses},
  JOURNAL =      {Science},
  YEAR =         {2009},
  volume =       {326},
  pages =        {1074-1077},
}

@ARTICLE{Hau99Nature,
  AUTHOR =       {L. V. Hau and S. E. Harris and Z. Dutton and C. Behroozi},
  TITLE =        {Light speed reduction to 17 m per second in an ultracold atomic gas},
  JOURNAL =      {Nature},
  YEAR =         {1999},
  volume =       {397},
  pages =        {594-598},
}

@ARTICLE{Kash99PRL,
  AUTHOR =       {M. M. Kash and V. A. Sautenkov and A. S. Zibrov and L. Hollberg and G. R. Welch and M. D. Lukin and Y. Rostovtsev and E. S. Fry and M. O. Scully},
  TITLE =        {Ultraslow group velocity and enhanced nonlinear optical effects in a coherently driven hot atomic gas},
  JOURNAL =      {Phys. Rev. Lett.},
  YEAR =         {1999},
  volume =       {82},
  pages =        {5229-5232},
}

@ARTICLE{Song05OE,
  AUTHOR =       {K. Y. Song and M. G. Herr{\'a}ez and L. Th{\'e}venaz},
  TITLE =        {Gain-assisted pulse advancement using single and double {B}rillouin gain peaks in optical fibers},
  JOURNAL =      {Opt. Express},
  YEAR =         {2005},
  volume =       {13},
  pages =        {9758-9765},
}

@ARTICLE{Baba08NP,
  AUTHOR =       {T. Baba},
  TITLE =        {Slow light in photonic crystals},
  JOURNAL =      {Nat. Photon.},
  YEAR =         {2008},
  volume =       {2},
  pages =        {465-473},
}

@ARTICLE{Tsakmakidis17Science,
  AUTHOR =       {K. L. Tsakmakidis and O. Hess and R. W. Boyd and X. Zhang},
  TITLE =        {Ultraslow waves on the nanoscale},
  JOURNAL =      {Science},
  YEAR =         {2017},
  volume =       {358},
  pages =        {eaan5196},
}

@ARTICLE{Giovannini15Science,
  AUTHOR =       {D. Giovannini and J. Romero and V. Poto{\v c} and G. Ferenczi and F. Speirits and S. M. Barnett and D. Faccio and M. J. Padgett},
  TITLE =        {Spatially structured photons that travel in free space slower than the speed of light},
  JOURNAL =      {Science},
  YEAR =         {2015},
  volume =       {347},
  pages =        {857-860},
}

@ARTICLE{Bouchard16Optica,
  AUTHOR =       {F. Bouchard and J. Harris and H. Mand and R. W. Boyd and E. Karimi},
  TITLE =        {Observation of subluminal twisted light in vacuum},
  JOURNAL =      {Optica},
  YEAR =         {2016},
  volume =       {3},
  pages =        {351-354},
}

@ARTICLE{Lyons18Optica,
  Title =        {How fast is a twisted photon?},
  AUTHOR =       {A. Lyons and T. Roger and N. Westerberg and S. Vezzoli and C. Maitland and J. Leach and M. J. Padgett and D. Faccio},
  JOURNAL =      {Optica},
  YEAR =         {2018},
  volume =       {5},
  pages =        {},
}

@ARTICLE{Froula18NP,
  AUTHOR =       {D. H. Froula and D. Turnbull and A. S. Davies and T. J. Kessler and D. Haberberger and J. P. Palastro and S.-W. Bahk and I. A. Begishev and R. Boni and S. Bucht and J. Katz and J. L. Shaw},
  TITLE =        {Spatiotemporal control of laser intensity},
  JOURNAL =      {Nat. Photon.},
  YEAR =         {2018},
  volume =       {12},
  pages =        {262-265},
}

@ARTICLE{Zapata06OL,
  AUTHOR =       {C. J. Zapata-Rodr{\'i}guez and M. A. Porras},
  TITLE =        {X-wave bullets with negative group velocity in vacuum},
  JOURNAL =      {Opt. Lett.},
  YEAR =         {2006},
  volume =       {31},
  pages =        {3532-3534},
}

@ARTICLE{Kuntz09PRA,
  AUTHOR =       {K. B. Kuntz and B. Braverman and S. H. Youn and M. Lobino and E. M. Pessina and A. I. Lvovsky},
  TITLE =        {Spatial and temporal characterization of a Bessel beam produced using a conical mirror},
  JOURNAL =      {Phys. Rev. A},
  YEAR =         {2009},
  volume =       {79},
  pages =        {043802},
}

@ARTICLE{Bowlan09OL,
  AUTHOR =       {P. Bowlan and H. Valtna-Lukner and M. L{\~o}hmus and P. Piksarv and P. Saari and R. Trebino},
  TITLE =        {Measuring the spatiotemporal field of ultrashort {B}essel-{X} pulses},
  JOURNAL =      {Opt. Lett.},
  YEAR =         {2009},
  volume =       {34},
  pages =        {2276-2278},
}

@ARTICLE{Bonaretti09OE,
  AUTHOR =       {F. Bonaretti and D. Faccio and M. Clerici and J. Biegert and P. {Di T}rapani},
  TITLE =        {Spatiotemporal amplitude and phase retrieval of {B}essel-{X} pulses using a {H}artmann-{S}hack sensor},
  JOURNAL =      {Opt. Express},
  YEAR =         {2009},
  volume =       {17},
  pages =        {9804-9809},
}

@ARTICLE{Liu98JMO,
  AUTHOR =       {Z. Liu and D. Fan},
  TITLE =        {Propagation of pulsed zeroth-order {B}essel beams},
  JOURNAL =      {J. Mod. Opt.},
  YEAR =         {1998},
  volume =       {45},
  pages =        {17-21},
}

@ARTICLE{Zapata08JOSAA,
  AUTHOR =       {C. J. Zapata-Rodr{\'i}guez and M. A. Porras and J. J. Miret},
  TITLE =        {Free-space delay lines and resonances with ultraslow pulsed {B}essel beams},
  JOURNAL =      {J. Opt. Soc. Am. A},
  YEAR =         {2008},
  volume =       {25},
  pages =        {2758-2763},
}

@ARTICLE{Averchi08PRA,
  AUTHOR =       {A. Averchi and D. Faccio and R. Berlasso and M. Kolesik and J. V. Moloney and A. Couairon and P. {Di T}rapani},
  TITLE =        {Phase matching with pulsed {B}essel beams for high-order harmonic generation},
  JOURNAL =      {Phys. Rev. A},
  YEAR =         {2008},
  volume =       {77},
  pages =        {021802(R)},
}

@ARTICLE{Danielius96OL,
  AUTHOR =       {R. Danielius and A. Piskarskas and P. {Di T}rapani and A. Andreoni and C. Solcia and P. Foggi},
  TITLE =        {Matching of group velocities by spatial walk-off in collinear three-wave interaction with tilted pulses},
  JOURNAL =      {Opt. Lett.},
  YEAR =         {1996},
  volume =       {21},
  pages =        {973-975},
}

@ARTICLE{Porras03PRE2,
  AUTHOR =       {M. A. Porras and G. Valiulis and P. {Di T}rapani},
  TITLE =        {Unified description of {B}essel {X} waves with cone dispersion and tilted pulses},
  JOURNAL =      {Phys. Rev. E},
  YEAR =         {2003},
  volume =       {68},
  pages =        {016613},
}

@ARTICLE{Porras03PRE,
  AUTHOR =       {M. A. Porras and I. Gonzalo and A. Mondello},
  TITLE =        {Pulsed light beams in vacuum with superluminal and negative group velocities},
  JOURNAL =      {Phys. Rev. E},
  YEAR =         {2003},
  volume =       {67},
  pages =        {066604},
}

@ARTICLE{Hebling02OE,
  AUTHOR =       {J. Hebling and G. Almási and I. Z. Kozma and J. Kuhl},
  TITLE =        {Velocity matching by pulse front tilting for large-area {TH}z-pulse generation},
  JOURNAL =      {Opt. Express},
  YEAR =         {2002},
  volume =       {10},
  pages =        {1161-1166},
}

@ARTICLE{Torres10AOP,
  AUTHOR =       {J. P. Torres and M. Hendrych and A. Valencia},
  TITLE =        {Angular dispersion: an enabling tool in nonlinear and quantum optics},
  JOURNAL =      {Adv. Opt. Photon.},
  YEAR =         {2010},
  volume =       {2},
  pages =        {319-369},
}

@INCOLLECTION{Fulop10Review,
  AUTHOR =       {J. A. F{\"u}l{\"o}p and J. Hebling},
  TITLE =        {Applications of tilted-pulse-front excitation},
  editor =       {K. Y. Kim},
  YEAR =         {2010},
  publisher =    {InTech},
  booktitle =    {Recent Optical and Photonic Technologies},
}

@ARTICLE{Hebling08JOSAB,
  AUTHOR =       {J. Hebling and K.-L. Yeh and M. C. Hoffmann and B. Bartal and K. A. Nelson},
  TITLE =        {Generation of high-power terahertz pulses by tilted-pulse-front excitation and their application possibilities},
  JOURNAL =      {J. Opt. Soc. Am. B},
  YEAR =         {2008},
  volume =       {25},
  pages =        {B6-B19},
}

@ARTICLE{Wang20LPR,
  AUTHOR =       {L. Wang and G. T{\'o}th and J. Hebling and F. K{\"a}rtner},
  TITLE =        {Tilted-pulse-front schemes for terahertz generation},
  JOURNAL =      {Laser Photon. Rev.},
  YEAR =         {2020},
  volume =       {14},
  pages =        {2000021},
}

@ARTICLE{Shabahang17SR,
  AUTHOR =       {S. Shabahang and H. E. Kondakci and M. L. Villinger and J. D. Perlstein and A. {El H}alawany and A. F. Abouraddy},
  TITLE =        {Omni-resonant optical micro-cavity},
  JOURNAL =      {Sci. Rep.},
  YEAR =         {2017},
  volume =       {7},
  pages =        {10336},
}

@ARTICLE{Yessenov19OE,
  AUTHOR =       {M. Yessenov and B. Bhaduri and L. Mach and D. Mardani and H. E. Kondakci and M. A. Alonso and G. A. Atia and A. F. Abouraddy},
  TITLE =        {What is the maximum differential group delay achievable by a space-time wave packet in free space?},
  JOURNAL =      {Opt. Express},
  YEAR =         {2019},
  volume =       {27},
  pages =        {12443-12457},
}

@ARTICLE{Yessenov20NC,
  AUTHOR =       {M. Yessenov and B. Bhaduri and P. J. Delfyett and A. F. Abouraddy},
  TITLE =        {Free-space optical delay line using space-time wave packets},
  JOURNAL =      {Nat. Commun.},
  YEAR =         {2020},
  volume =       {11},
  pages =        {5782},
}

@ARTICLE{Hall21OL,
  AUTHOR =       {L. A. Hall and M. Yessenov and A. F. Abouraddy},
 title={Space--time wave packets violate the universal relationship between angular dispersion and pulse-front tilt},
  journal={Opt. Lett.},
  volume={46},
  number={7},
  pages={1672--1675},
  year={2021},
  publisher={Optical Society of America}
}

@Article{Hall21APLSTTalbot,
  author  = {L. A. Hall and M. Yessenov and S. A. Ponomarenko and A. F. Abouraddy},
  journal = {APL Photon.},
  title   = {The space-time {T}albot effect},
  year    = {2021},
  pages   = {056105},
  volume  = {6},
}

@article{Yessenov22AOP,
    title={Space-time wave packets},
    author={Yessenov, M. and Hall, L. A. and Schepler, K. L. and Abouraddy, A. F.},
    journal = {Adv. Opt. Photon.},
    pages = {455-570},
    volume = {14},
    year = {2022},
}

@article{Yessenov22NC,
  title={Space-time wave packets localized in all dimensions},
  author={M. Yessenov and J. Free and Z. Chen and E. G. Johnson and M. P. J. Lavery and M. A. Alonso and A. F. Abouraddy},
  journal={Nat. Commun.},
  volume={13},
  pages={4573},
  year={2022},
}

@ARTICLE{Yessenov22OL,
  AUTHOR =       {M. Yessenov and Z. Chen and M. P. J. Lavery and A. F. Abouraddy},
  TITLE =        {Vector space-time wave packets},
  JOURNAL =      {Opt. Lett.},
  YEAR =         {2022},
  volume =       {47},
  pages =        {4131-4134},
}

@ARTICLE{Shiri20OL,
  AUTHOR =       {A. Shiri and M. Yessenov and R. Aravindakshan and A. F. Abouraddy},
  TITLE =        {Omni-resonant space-time wave packets},
  JOURNAL =      {Opt. Lett.},
  YEAR =         {2020},
  volume =       {45},
  pages =        {},
}

@ARTICLE{Hall22OEConsequences,
  AUTHOR =       {L. A. Hall and A. F. Abouraddy},
  TITLE =        {Consequences of non-differentiable angular dispersion in optics: tilted pulse fronts versus space-time wave packets},
  JOURNAL =      {Opt. Express},
  YEAR =         {2022},
  volume =       {30},
  pages =        {4817-4832},
}

@ARTICLE{Hall22JOSAA,
  AUTHOR =       {L. A. Hall and A. F. Abouraddy},
  TITLE =        {Non-differentiable angular dispersion as an optical resource},
  JOURNAL =      {J. Opt. Soc. Am. A},
  YEAR =         {2022},
  volume =       {39},
  pages =        {2016-2025},
}

@ARTICLE{Mhibik23OL,
  AUTHOR =       {O. Mhibik and M. Yessenov and L. Mach and L. Glebov and A. F. Abouraddy and I. Divliansky},
  TITLE =        {Rotated chirped volume Bragg gratings for compact spectral analysis},
  JOURNAL =      {Opt. Lett.},
  YEAR =         {2023},
  volume =       {48},
  pages =        {1180-1183},
}

@ARTICLE{Yessenov23OL,
  AUTHOR =       {M. Yessenov and O. Mhibik and L. Mach and T. M. Hayward and R. Menon and L. Glebov and I. Divliansky and A. F. Abouraddy},
  TITLE =        {Ultracompact system for synthesizing space-time wave packets},
  JOURNAL =      {Opt. Lett.},
  YEAR =         {2023},
  volume =       {48},
  pages =        {2500-2503},
}

@ARTICLE{Mhibik23OL2,
  AUTHOR =       {O. Mhibik and M. Yessenov and L. Glebov and A. F. Abouraddy and I. Divliansky},
  TITLE =        {Compact dual-band spectral analysis via multiplexed rotated chirped volume Bragg gratings},
  JOURNAL =      {Opt. Lett.},
  YEAR =         {2023},
  volume =       {48},
  pages =        {5137-5140},
}

@ARTICLE{Shen23JO,
  AUTHOR =       {Y. Shen and Q. Zhan and L. G. Wright and D. N. Christodoulides and F. W. Wise and A. E. Willner and Z. Zhao and K. Zou and C.-T. Liao and C. Hern{\'a}ndez-Garc{\'i}a and M. Murnane and M. A. Porras and A. Chong and C. Wan and K. Y. Bliokh and M. Yessenov and A. F. Abouraddy and L. J. Wong and M. Go and S. Kumar and C. Guo and S. Fan and N. Papasimakis and N. I. Zheludev and L. Chen and W. Zhu and A. Agrawal and S. W. Jolly and C. Dorrer and B. Alonso and I. Lopez-Quintas and M. L{\'o}pez-Ripa and {\'I}. J. Sola and Y. Fang and Q. Gong and Y. Liu and J. Huang and H. Zhang and Z. Ruan and M. Mounaix and N. K. Fontaine and J. Carpenter and A. H. Dorrah and F. Capasso and A. Forbes},
  TITLE =        {Roadmap on spatiotemporal light fields},
  JOURNAL =      {J. Opt.},
  YEAR =         {2023},
  volume =       {25},
  pages =        {093001},
}

@ARTICLE{Hall21OLNormalGVD,
  AUTHOR =       {L. A. Hall and A. F. Abouraddy},
  TITLE =        {Realizing normal group-velocity dispersion in free space via angular dispersion},
  JOURNAL =      {Opt. Lett.},
  YEAR =         {2021},
  volume =       {46},
  pages =        {5421-5424},
}

@ARTICLE{Hall21PRAVwave,
  AUTHOR =       {L. A. Hall and A. F. Abouraddy},
  TITLE =        {Free-space group-velocity dispersion induced in space-time wave packets by {V}-shaped spectra},
  JOURNAL =      {Phys. Rev. A},
  YEAR =         {2021},
  volume =       {104},
  pages =        {013505},
}

@ARTICLE{Hall24JOSAA,
  AUTHOR =       {L. A. Hall and A. F. Abouraddy},
  TITLE =        {Universal angular-dispersion synthesizer},
  JOURNAL =      {J. Opt. Soc. Am. A},
  YEAR =         {2024},
  volume =       {41},
  pages =        {83-94},
}

@ARTICLE{Hall25LRR,
  AUTHOR =       {L. A. Hall and A. Shiri and A. F. Abouraddy},
  TITLE =        {Omni-resonant imaging across the visible},
  JOURNAL =      {Laser Photon. Rev.},
  YEAR =         {2025},
  volume =       {19},
  pages =        {2500270},
}

@ARTICLE{Piccardo23NP,
  AUTHOR =       {M. Piccardo and M. {de O}liveira and V. R. Policht and M. Russo and B. Ardini and M. Corti and G. Valentini and J. Vieira and C. Manzoni and G. Cerullo and A. Ambrosio},
  TITLE =        {Broadband control of topological–spectral correlations in space–time beams},
  JOURNAL =      {Nat. Photon.},
  YEAR =         {2023},
  volume =       {17},
  pages =        {822-828},
}

@ARTICLE{Romer25JOpt,
  AUTHOR =       {M. A. Romer and L. A. Hall and A. F. Abouraddy},
  TITLE =        {Synthesis and characterization of space-time light sheets: a tutorial},
  JOURNAL =      {J. Opt.},
  year =         {2025},
  volume =       {27},
  pages =        {013501},
}

@ARTICLE{Abouraddy25OPN,
  AUTHOR =       {A. F. Abouraddy and M. Yessenov and I. Divliansky and A. T. Watnik},
  TITLE =        {New frontiers in spatiotemporally structured light},
  JOURNAL =      {Opt. Photon. News},
  YEAR =         {2025},
  volume =       {36},
  pages =        {38-45},
}

@ARTICLE{Sambles15Science,
  AUTHOR =       {J. R. Sambles},
  TITLE =        {Structured photons take it slow},
  JOURNAL =      {Science},
  YEAR =         {2015},
  volume =       {347},
  pages =        {828},
}

@ARTICLE{Saari17Optica,
  AUTHOR =       {P. Saari},
  TITLE =        {Observation of subluminal twisted light in vacuum: comment},
  JOURNAL =      {Optica},
  YEAR =         {2017},
  volume =       {4},
  pages =        {204-206},
}

@ARTICLE{Hall25APLP,
  AUTHOR =       {L. A. Hall and M. Yessenov and K. L. Schepler and A. F. Abouraddy},
  TITLE =        {Universality and Non-differentiability: {A} new perspective on angular dispersion in optics},
  JOURNAL =      {APL Photon.},
  YEAR =         {2025},
  volume =       {10},
  pages =        {121101},
}

@article{Vigneron10PRE,
	author =	{J. P. Vigneron and P. Simonis and A. Aiello and A. Bay and D. M. Windsor and J.-F. Colomer and M. Rassart},
	journal =	{Phys. Rev. E},
	pages =		{021903},
	title =		{Reverse color sequence in the diffraction of white light by the wing of the male butterfly \textit{Pierella luna} (Nymphalidae: Satyrinae)},
	volume =	{82},
	year =		{2010}
}

@article{England14PNAS,
	author =	{G. England and M. Kolle and P. Kim and M. Khan and P. Mu{\~n}oz and E. Mazur and J. Aizenberg},
	journal =	{Proc. Natl. Acad. Sci. USA},
	pages =		{15630-15634},
	title =		{Bioinspired micrograting arrays mimicking the reverse color diffraction elements evolved by the butterfly \textit{Pierella luna}},
	volume =	{111},
	year =		{2014}
}

@article{Bartl4PNAS,
	author =	{M. H. Bartl},
	journal =	{Proc. Natl. Acad. Sci. USA},
	pages =		{15602-15603},
	title =		{Butterfly-inspired photonic reverse diffraction color sequence},
	volume =	{111},
	year =		{2014}
}

@article{Arbabi2017Optica,
  title={Controlling the sign of chromatic dispersion in diffractive optics with dielectric metasurfaces},
  author={E. Arbabi and A. Arbabi and S. M. Kamali and Y. Horie and A. Faraon},
  journal={Optica},
  volume={4},
  pages={625-632},
  year={2017},
}

@article{McClung2020Light,
  title={At-will chromatic dispersion by prescribing light trajectories with cascaded metasurfaces},
  author={A. McClung and M. Mansouree and A. Arbabi},
  journal={Light Sci. Appl.},
  volume={9},
  pages={93},
  year={2020},
}

@article{Kamali2018Review,
  title={A review of dielectric optical metasurfaces for wavefront control},
  author={S. M. Kamali and E. Arbabi and A. Arbabi and A. Faraon},
  journal={Nanophoton.},
  volume={7},
  pages={1041-1068},
  year={2018},
}

@article{Li2021SciAdv,
  title={Meta-optics achieves {RGB}-achromatic focusing for virtual reality},
  author={Li, Z. and Lin, P. and Huang, Y. and Park, J. and Chen, W. T. and Shi, Z. and Qiu, Cheng. and Cheng, J. and Capasso, F.},
  journal={Sci. Adv.},
  volume={7},
  pages={eabe4458},
  year={2021},
}

@article{Wang2018Science,
  title={A broadband achromatic metalens in the visible},
  author={S. Wang and P. C. Wu and V.-C. Su and Y.-C. Lai and M.-K. Chen and H. Y. Kuo and B. H. Chen and Y. H. Chen and T.-T. Huang and J.-H. Wang and R.-M. Lin and C.-H. Kuan and T. Li and Z. Wang and S. Zhu and D. P. Tsai},
  journal={Nat. Nanotech.},
  volume={13},
  pages={227-232},
  year={2018},
}

@ARTICLE{Zhang20LSA,
  AUTHOR =       {X. Zhang and Q. Li and F. Liu and M. Qiu and S. Sun and Q. He and L. Zhou},
  TITLE =        {Controlling angular dispersions in optical metasurfaces},
  JOURNAL =      {Light Sci. Appl.},
  YEAR =         {2020},
  volume =       {9},
  pages =        {76},
}

@ARTICLE{Qiu18PRAppl,
  AUTHOR =       {M. Qiu and M. Jia and S. Ma and S. Sun and Q. He and L. Zhou},
  TITLE =        {Angular dispersions in terahertz metasurfaces: {P}hysics and applications},
  JOURNAL =      {Phys. Rev. Applied},
  YEAR =         {2018},
  volume =       {9},
  pages =        {054050},
}

@ARTICLE{Parra07OPN,
  AUTHOR =       {E. Parra and J. R. Lowell},
  TITLE =        {Toward applications of slow light technology},
  JOURNAL =      {Opt. Photon. News},
  YEAR =         {2007},
  volume =       {18},
  pages =        {40-45},
}

@ARTICLE{Krauss07JPD,
  AUTHOR =       {T. F. Krauss},
  TITLE =        {Slow light in photonic crystal waveguides},
  JOURNAL =      {J. Phys. D},
  YEAR =         {2007},
  volume =       {40},
  pages =        {2666-2670},
}

@ARTICLE{Krauss08NP,
  AUTHOR =       {T. F. Krauss},
  TITLE =        {Why do we need slow light?},
  JOURNAL =      {Nat. Photon.},
  YEAR =         {2008},
  volume =       {2},
  pages =        {448-450},
}

@BOOK{Kurgin08Book,
  Editor =       {J. B. Khurgin and R. S. Tucker},
  TITLE =        {Slow Light: {S}cience and Applications},
  YEAR =         {2008},
  publisher =    {CRC Press},
  address =      {Boca Raton, FL},
}

@ARTICLE{Tucker05EL,
  AUTHOR =       {R. S. Tucker and P. C. Ku and C. J. Chang-Hasnain},
  TITLE =        {Delay-bandwidth product and storage density in slow-light optical buffers},
  JOURNAL =      {Electron. Lett.},
  YEAR =         {2005},
  volume =       {41},
  pages =        {208-209},
}

@ARTICLE{Tucker05JLT,
  AUTHOR =       {R. S. Tucker and P. C. Ku and C. J. Chang-Hasnain},
  TITLE =        {Slow-light optical buffers: {c}apabilities and fundamental limitations},
  JOURNAL =      {{J. Lightwave} Technol.},
  YEAR =         {2005},
  volume =       {23},
  pages =        {4046-4066},
}

@ARTICLE{Vlasov05N,
  AUTHOR =       {Y. A. Vlasov and M. {O`B}oyle and H. F. Hamann and S. J. McNab},
  TITLE =        {Active control of slow light on a chip with photonic crystal waveguides},
  JOURNAL =      {Nature},
  YEAR =         {2005},
  volume =       {438},
  pages =        {65-69},
}

@ARTICLE{Okawachi05PRL,
  AUTHOR =       {Y. Okawachi and M. S. Bigelow and J. E. Sharping and Z. Zhu and A. Schweinsberg and D. J. Gauthier and R. W. Boyd and A. L. Gaeta},
  TITLE =        {Tunable all-optical delays via {B}rillouin slow light in an optical fiber},
  JOURNAL =      {Phys. Rev. Lett.},
  YEAR =         {2005},
  volume =       {94},
  pages =        {153902},
}

@ARTICLE{Yanik04PRL,
  AUTHOR =       {M. F. Yanik and S. Fan},
  TITLE =        {Stopping light all optically},
  JOURNAL =      {Phys. Rev. Lett.},
  YEAR =         {2004},
  volume =       {92},
  pages =        {083901},
}

@ARTICLE{Xu06NPhys,
  AUTHOR =       {Q. Xu and P. Dong and M. Lipson},
  TITLE =        {Breaking the delay-bandwidth limit in a photonic structure},
  JOURNAL =      {Nat. Phys.},
  YEAR =         {2007},
  volume =       {3},
  pages =        {406-410},
}

@ARTICLE{Tsakmakidis17Science2,
  AUTHOR =       {K. L. Tsakmakidis and L. Shen and S. A. Schulz and X. Zheng and J. Upham and X. Deng and H. Altug and A. F. Vakakis and R. W. Boyd},
  TITLE =        {Breaking {L}orentz reciprocity to overcome the time-bandwidth limit in physics and engineering},
  JOURNAL =      {Science},
  YEAR =         {2017},
  volume =       {356},
  pages =        {1260-1264},
}

@ARTICLE{Martinez84JOSAA,
  AUTHOR =       {O. E. Martinez and J. P. Gordon and R. L. Fork},
  TITLE =        {Negative group-velocity dispersion using refraction},
  JOURNAL =      {J. Opt. Soc. Am. A},
  YEAR =         {1984},
  volume =       {1},
  pages =        {1003-1006},
}

@ARTICLE{Yessenov25JOSAA,
  AUTHOR =       {M. Yessenov and A. F. Abouraddy},
  TITLE =        {Optical spatiotemporal {F}ourier synthesis: tutorial},
  JOURNAL =      {J. Opt. Soc. Am. A},
  YEAR =         {2025},
  volume =       {42},
  pages =        {1295-1315},
}

@article{Vieira18PRL,
    author = {J. Vieira and J. T. Mendon{\v c}a and F. Qu{\'e}r{\'e}},
    title = {Optical control of the topology of laser-plasma accelerators},
    journal = {Phys. Rev. Lett.},
    year = {2018},
    volume = {121},
    pages = {054801},
}

@article{Palastro20PRL,
  author  = {J. P. Palastro and J. L. Shaw and P. Franke and D. Ramsey
             and T. T. Simpson and D. H. Froula},
  title   = {Dephasingless laser wakefield acceleration},
  journal = {Phys. Rev. Lett.},
  volume  = {124},
  pages   = {134802},
  year    = {2020},
}

@article{Sun24PRR,
    author = {F. Sun and W. Wang and H. Dong and J. He and Z. Shi and Z. Lv and Q. Zhan and Y. Leng and S. Zhuang and R. Li},
    title = {Generation of isolated attosecond electron sheet via relativistic spatiotemporal optical manipulation},
    journal = {Phys. Rev. Research},
    year = {2024},
    volume = {6},
    pages = {013075},
}

@article{Piccardo25Optica,
    author = {M. Piccardo and M. O. Cernaianu and J. P. Palastro and A. Arefiev and C. Thaury and J. Vieira and D. H. Froula and V. Malka},
    title = {Trends in relativistic laser–matter interaction: the promises of structured light},
    journal = {Optica},
    year = {2025},
    volume = {12},
    pages = {732-752},
}

@article{Longman26NP,
    author = {A. Longman and D. Attiyah and E. S. Grace and C. Gardner and T. Suratwala and G. Tham and C. Harthcock and R. Fedosejevs and F. Dollar},
    title = {Spatiotemporal shaping of broadband helical light pulses at relativistic intensities},
    journal = {Nat. Photon.},
    year = {2026},
    volume = {20},
    pages = {843-849},
}

@article{Vaz26arxiv,
    author = {G. Vaz and R. Almeida and P. {San Miguel C}laveria and R. Neumann and J. Pereira and C. Miranda and V. Ginis and J. Vieira and M. Fajardo and M. Piccardo},
    title = {Space-time beams with tunable orbital group velocity for plasma superradiance},
    journal = {arXiv:2602.18838},
    year = {2026},
    volume = {},
    pages = {},
}

@article{Caizergues20NP,
	title = {Phase-locked laser-wakefield electron acceleration},
	volume = {14},
	journal = {Nat. Photon.},
	author = {Caizergues, C. and Smartsev, S. and Malka, V. and Thaury, C.},
	year = {2020},
	pages = {475--479},
}

@ARTICLE{Campbell90JASA,
  AUTHOR =       {J. A. Campbell and S. Soloway},
  TITLE =        {Generation of a nondiffracting beam with frequency-independent beamwidth},
  JOURNAL =      {J. Acoust. Soc. Am.},
  YEAR =         {1990},
  volume =       {88},
  pages =        {2467-2477},
}

@Article{Alfano2016OC,
  author  = {R. R. Alfano and D. A.Nolan},
  journal = {Opt. Commun.},
  title   = {Slowing of {B}essel light beam group velocity},
  year    = {2016},
  pages   = {25-27},
  volume  = {361},
}

@Article{Saari2017OC,
  author  = {P. Saari},
  journal = {Opt. Commun.},
  title   = {Comments on ``{S}lowing of {B}essel light beam group velocity''},
  year    = {2017},
  pages   = {300-301},
  volume  = {392},
}

@Article{Stenner2003Nature,
  author  = {M. D. Stenner and D. J. Gauthier and M. A. Neifeld},
  journal = {Nature},
  title   = {The speed of information in a `fast-light' optical medium},
  year    = {2003},
  pages   = {695-698},
  volume  = {425},
}

@ARTICLE{Landauer93Nature,
  AUTHOR =       {R. Landauer},
  TITLE =        {Light faster than light?},
  JOURNAL =      {Nature},
  YEAR =         {1993},
  volume =       {365},
  pages =        {692–693},
}

@ARTICLE{Landauer92SSC,
  AUTHOR =       {R. Landauer and Th. Martin},
  TITLE =        {Time delay in wave packet tunneling},
  JOURNAL =      {Solid State Commun.},
  YEAR =         {1992},
  volume =       {84},
  pages =        {115-117},
}

@ARTICLE{Landauer89Nature,
  AUTHOR =       {R. Landauer},
  TITLE =        {Barrier traversal time},
  JOURNAL =      {Nature},
  YEAR =         {1989},
  volume =       {341},
  pages =        {567-568},
}

@ARTICLE{Diener96PLA,
  AUTHOR =       {G. Diener},
  TITLE =        {Superluminal group velocities and information transfer},
  JOURNAL =      {Phys. Lett. A},
  YEAR =         {1996},
  volume =       {223},
  pages =        {327-331},
}

@ARTICLE{Milonni02JPB,
  AUTHOR =       {P. W Milonni},
  TITLE =        {Controlling the speed of light pulses},
  JOURNAL =      {J. Phys. B},
  YEAR =         {2002},
  volume =       {35},
  pages =        {R31-R56},
}

@Book{Milonni20Book,
  AUTHOR =       {P. W. Milonni},
  TITLE =        {Fast Light, Slow Light and Left-Handed Light},
  Publisher =    {CRC Press},
  YEAR =         {2020},
}

@ARTICLE{Chiao02OPN,
  AUTHOR =       {R. Y. Chiao and P. W. Milonni},
  TITLE =        {Fast light, slow light},
  JOURNAL =      {Opt. Photon. News},
  YEAR =         {2002},
  issue =        {6},
  volume =       {13},
  pages =        {26-30},
}

@ARTICLE{Smith70AJP,
  AUTHOR =       {R. L. Smith},
  TITLE =        {The velocities of light},
  JOURNAL =      {Am. J. Phys.},
  YEAR =         {1970},
  volume =       {38},
  pages =        {978-984},
}

@ARTICLE{Hall26JOSAA2,
  AUTHOR =       {L. A. Hall and A. F. Abouraddy},
  TITLE =        {Limits on the free-space group velocity of optical wave packets incorporating angular dispersion. {Part~II}, non-differentiable angular dispersion: tutorial},
  JOURNAL =      {accompanying paper},
  YEAR =         {2026},
  volume =       {},
  pages =        {},
}

\end{document}